\documentclass[11pt, twocolumn]{article}
\usepackage[english]{babel}
\usepackage{graphicx,subcaption}
\usepackage{mathptm}
\usepackage{newtxtext,newtxmath}
\usepackage{amsmath,abstract} 
\usepackage{calrsfs}
\usepackage{physics}
\usepackage[
    backend=biber,
    style=numeric-comp,
    maxcitenames=2,
    maxbibnames=4,
    backref=true,
    sorting=none,
]{biblatex}
\usepackage{hyperref}
\usepackage[affil-it]{authblk}
\usepackage{cleveref}
\usepackage{booktabs}
\usepackage{url}
\usepackage{dblfloatfix}
\usepackage[toc,page]{appendix}

\DeclareMathAlphabet{\pazocal}{OMS}{zplm}{m}{n}

\newcommand{\bk}{\boldsymbol{k}}
\newcommand{\p}{\boldsymbol{ p}}

\newcommand{\bv}{\boldsymbol{ v}}

\newcommand{\F}{\boldsymbol{ F}}

\newcommand{\bK}{\boldsymbol{ K}}

\newcommand{\Q}{\boldsymbol{ Q}}

\usepackage[parfill]{parskip}

\begin{document}
\title{Climate Model Tuning with \\ Adaptive Supermodeling }

\author[1,*]{Jordan Seneca}
\author[2]{Suzanne Bintanja}
\author[1]{Frank M. Selten}
\affil[1]{Koninklijk Nederlands Meteorologisch Instituut, De Bilt, 3730 AE, Netherlands}
\affil[2]{Mani L. Bhaumik Institute for Theoretical Physics, University of California, Los Angeles, California, CA 90095, USA}
\affil[*]{Corresponding author: Jordan Seneca, \url{jordan.seneca@knmi.nl}}

\maketitle
\begin{abstract}

In climate science, the tuning of climate models is a computationally intensive problem due to the combination of the high-dimensionality of the system state and long integration times. Supermodelling is a technique which has shown the potential for reducing climate model biases by dynamically coupling multiple models together, and training their coupling on a short timescale. Here, we introduce a new approach called \emph{adaptive supermodeling}, where the internal model parameters of the member of a supermodel are tuned. We perform three experiments. We first directly optimize the internal parameters of a climate model. We then optimize the weights between two members of a supermodel in a classical supermodel approach. For a case designed to challenge the two previous methods, we implement adaptive supermodeling, which achieves a performance similar to a perfect model.


\end{abstract}

\section{Introduction}

Climate models are essential tools for studying the Earth system and making projections about the future climate. However, significant uncertainties in climate simulations exist, which are due to model structure, parameter uncertainty, internal variability, and uncertainties in external forcings. Two important avenues to reduce these uncertainties are to improve existing models by tuning or involving deep learning methods, or to combine multiple models together. 

First, models contain numerous parameters whose values are uncertain. For one, the value of effective parameters cannot be directly measured. Additionally, the optimal value of certain parameters depends on structural errors in parameterizations or errors in other parts of the model. These parameters are therefore routinely tuned to improve consistency with observed climatology \cite{modeltuning1,modeltuning2,ecearth_tuning3}. Due to the long adjustment timescales of the climate system, climate model tunings must be tested on long timescales. This makes algorithmic solutions expensive to implement, and tuning of climate models is therefore still often mainly done by experts manually selecting and testing sets of parameters. New, efficient algorithms for parameter tuning of weather models are therefore of interest to the climate modeling community. Recently, renewed effort to reduce climate model biases has brought forth the proposal of new methods \cite{derivative-free, oil_tuning}, including modern machine learning approaches \cite{MIT_nudged,RAIN}. As the development of weather and climate models has become increasingly merged in the past years, tuning on weather timescales can also offer an avenue to improved climate projections.

Weather and climate models have historically been developed largely independently, specializing in regional predictions on the scale of days to months, and global predictions on the scale of decades and above, respectively. In recent years however, climate change has spurred the need for climate information on a regional, seasonal and decadal scale \cite{TwoTimeScalesforThePriceOfOneAlmost}. This has motivated the use of high-resolution weather models on much longer timescales, as well as the increase of climate model resolutions for regional use. The latter being anyway desirable, as it makes climate models less reliant on potentially biased parameterizations of physical processes on large scales by resolving them instead.

Increasingly available computational power has made the overlap of these two classes of models a reality, and has resulted in an increased merged approach towards development of weather-to-climate predictions. Examples of this are seamless forecasting \cite{seamless,seamless2}, the EC-Earth3 climate model, whose atmospheric component is a modified version of ECMWF's Integrated Forecasting System (IFS) developed for medium-range weather forecasting, or the Deutscher Wetterdienst's ICON model \cite{ICON} which spans predictions from regional to global, and weather to climate time scales. Additionally, climate dynamics can in some cases be captured on relatively short timescales of a few years \cite{climate_variability}. This means that improvements of weather models on this timescale has the benefit of also potentially improving long-term climate projections made by the same model. 

Similarly to climate models, one way of improving weather models is to tune their parameters. Contrary to climate models however, algorithmic tuning schemes for weather models are computationally feasible, and have been implemented for the IFS, for example the Ensemble Prediction and Parameter Estimation System (EPPES) \cite{skill_optimization_ECMWF_2013,eppes}. Another approach is the use of deep learning with reanalysis or observational data, which has brought forth an era of machine learning (ML) models which have shown to be capable of outperforming conventional numerical weather prediction (NWP) models in standard metrics such as the RMSE of medium-range forecasting \cite{aifs, graphcast, subseasonal_ml, gencast}. Despite the superior performance of ML models in some metrics, they still struggle in predicting extremes or capturing certain atmospheric phenomena as well as NWP models \cite{comparison_ai_forecasting, aifs_ifs_comparison}. Since ML and NWP models appear to both have their own advantages and drawbacks, the development of hybrid methods, which combine the strengths of both, has become increasingly popular \cite{hybrid_noaa, hu2026hybrid, hybrid_india}.

Second, given different choices made in building different climate models, another option to improve climate model predictions is to combine simulations made by different models, reducing individual model biases. Multi-model ensemble projections, such as the Coupled Model Intercomparison Project (CMIP) \cite{cmip6}, or the North American Multi-Model Ensemble (NMME) \cite{NMME}, successfully use this approach albeit with some persistence of biases in extremes \cite{cmip6_eval}. More accurate predictions could be achieved by multi-model ensembles with more sophisticated ensemble weighting schemes \cite{non_intearcting_ensemble}. However, this type of  climate model combination remains limited since biases are only reduced by averaging in post-processing, missing out on any improvement on the dynamical behavior that could arise from the reduced biases.

\emph{Supermodeling} (SUMO) is a method which combines multiple models into a single dynamical system \cite{Supermodeling}. Unlike conventional multi-model ensembles, the members of a SUMO influence one another during integration, and produce a single system which evolves informed by the physics of the member models. A SUMO is effectively a single model evolving dynamically while also benefiting from the bias canceling of multi-model ensembles. SUMO is also an option for creating a hybrid model where the members can be both NWP and ML models.

The members of a SUMO can either be combined by dynamic coupling (connected SUMO), or by a weighted average of their tendencies (weighted SUMO) \cite{schevenhoven2021training,wiegerinck2013limit}. With an appropriate choice of coupling strengths or weights, the bias of the SUMO can be corrected. Previously, a connected SUMO was used \cite{selten2017simulating} with the SPEEDO climate model \cite{severijns2010efficient}. In this work, a weighted SUMO is used, where the state of each model is set to be identical, and the tendency is set to a weighted sum of the tendencies of each member. The weights of the tendencies are then optimized. So far, only the connection coefficients or tendency weights of a SUMO has been tested, but not the internal parameters of member models.

In this paper, we propose a new method called \emph{adaptive supermodeling} (ASUMO). The idea of ASUMO is to implement weather model-style parameter tuning in a SUMO for the purposes of improving the long term behaviour of the model. In ASUMO, both the weights and the internal parameters of one of the member models are optimized. This method gives more flexibility to the SUMO, and permits the direct tuning of member models of a SUMO. This expands the horizon of the search for an optimal set of parameters, also known as the \emph{convex hull} \cite{schevenhoven2017efficient}.

We train both the SUMO weights and internal parameters using a simple gradient descent method which estimates parameter values in the context of synchronization of chaos \cite{yuwu1996,duane2007identical,ospe025}. This method is based on the ansatz that, for an appropriate coupling strength between a perfect model and reference data, the model state will synchronize with the observations, resulting in a perfect representation of the truth. By additionally implementing a coupling to the model parameters, synchronization can be used on a non-perfect model to also adjust parameters to fit the data, achieving fast convergence and thus efficient parameter optimization. The algorithm operates sequentially on a time series, meaning that it can operate as data come in (online). Online tuning of model parameters is notably used by the Deutscher Wetterdienst's operational forecasting system \cite{online_wetterdienst}.

To demonstrate the potential of ASUMO for application in atmospheric science, we use the 3-level quasi-geostrophic global model of Marshall and Molteni \cite{marshall1993toward} at truncation 21 as the representation of both the truth and computational models. The parameters of the latter are given different values from those of the truth model, and are therefore referred as the \emph{imperfect} models. We present three different methods, all applied to a dataset of 1500 days. We first directly optimize the parameters of an imperfect model using a gradient descent method. We then apply the same method to the weights of a 2-member weighted SUMO of two imperfect models. Finally, we apply ASUMO to a configuration which is designed to be adversarial to both SUMO and training of parameters by themselves, demonstrating the expanded utility of ASUMO. 

We start in \cref{sec:methodology} by recalling the proof of the gradient descent parameter estimation method for chaotic systems. We then briefly describe the climate model used in this work (a more complete derivation is given in \cref{sec:qg}). Our configuration of the simulation experiments is also explained here. We then tune 6 parameters of a single model in \cref{sec:opt_parameters}, the weights of a SUMO in \cref{sec:sumo}, and both the weights of a SUMO and the parameters of one of its member models in \cref{sec:asumo}. For each case, we first  derive the update rule, followed by an experiment where we train the model using the method, and finally compare the resulting climatological errors to those of the perfect model. We finish with a discussion of the results in \cref{sec:discussion}. In \cref{sec:robustness}, we show a test of the robustness of the method to starting conditions, in \cref{sec:cost}, and estimate of thir computational costs, and in \cref{sec:adam}, we present the implementation of the adaptive learning rule algorithm. 



\section{Methodology} \label{sec:methodology}
\subsection{Parameter Estimation with Synchronization}\label{sec:ospe}

Duane, Yu, and Kocarev showed that synchronization enables parameter estimation in chaotic systems, and presented a simple gradient descent update rule to achieve this \cite{duane2007identical}. This method will be the basis for the optimization of both internal parameters and SUMO weights in this paper. Consider two coupled dynamical systems:

\begin{align}
\label{eq:cODEs}
\dot{\Q}^0&= \F(\Q^0;\p^0)  \nonumber \\
\dot{\Q}&= \F(\Q;\p) - \bK(\Q-\Q^0),  
\end{align}

where $\Q^0$ and $\Q$ are vectors of the system state, $\p^0$ and $\p$ are vectors of parameters, $K(\Q-\Q^0)$ is a nudging term that couples the two systems, and $K$ is the nudging strength. Suppose that when the two model parameters are identical, $\p = \p^0$, the two systems synchronize,  $ \lim_{t \rightarrow \infty} \mathbf{e}(t) = 0$, where $\mathbf{e}\equiv \Q-\Q^0$ is the state error (or \emph{ synchronization error}). Then, a Lyapunov function $L_0(\mathbf{e},\p)$ must exist for the case where $\p = \p^0$, such that $\dot{L}_0(\mathbf{e},\p^0)<0$ when $\mathbf{e}(t) \neq 0$ and 0 otherwise, for all $t > t_0$ for some time $t_0$ where synchronization is assumed to proceed monotonically. By adding a dynamical equation $\dot{p}=U$, with a well chosen update rule $U$, synchronization of both the system state and parameters can still occur, even if the two systems are not identical to begin with ($\p \ne \p^0$).

Consider the following update rule,
\begin{equation}
\label{eq:osperule}
 U \equiv  -r \sum_i \frac{\partial L_0}{\partial e_i} \frac {\partial h_i}{\partial p_j},
 \end{equation} 
where $h_i\equiv F_i(\Q;\p) -  F_i(\Q^0; \p^0)$ is the tendency error at index $i$, and $r$ is the learning rate which scales the magnitude of the parameter updating. If $\F$ is linear in $\p$, or if the error in $\p$ is small, i.e. $|\p - \p^0| \ll |\p^0|$, $L(\mathbf{e},\p\neq\p^0)$ is also a Lyapunov function, resulting in state synchronization as before, $\mathbf{e}\to 0$. In chaotic systems, this implies parameter estimation, $\p(t) \rightarrow \p^0$ \cite{comment_synch}. 

The resulting update rule is a familiar gradient descent learning rule which is equivalent to the delta rule used in the ADALINE algorithm by Widrow and Hoff \cite{widrow2025adaline}. The delta rule is widely used to train the weights of artificial neurons in single-layer neural network. In the present work, we will use this rule to directly train the parameters of a climate model instead. The choice of learning rate is bypassed by instead implementing the Adam algorithm which automatically scales updates to obtain fast and stable convergence. We additionally scale the updates by the current value of the parameter to start learning on scales relative to the parameter itself. The parameters are therefore updated by
\begin{equation}
    \dot{p}_j = p_j\text{Adam}(U),
\end{equation}
and $r\equiv1$ for all parameters. See \cref{sec:adam} for more details.

\subsection{ A 3-level Quasi-Geostrophic Atmospheric Model} 
The quasi-geostrophic global atmospheric model of Marshall and Molteni is used for this study \cite{marshall1993toward}. It reproduces large-scale physical features of the Earth's atmosphere with some realism, such as the jet streams, storms, and the short-term chaotic memory loss on the scale of a couple of weeks, while remaining computationally efficient. The model has three pressure levels, uses truncation number 21 in the spectral representation of its fields, 40 min timesteps, and potential vorticity as its only state variable. Six parameters from the model are investigated in this paper: the Ekman dissipation ($\tau_E^{-1}$), friction over land ($\alpha_1$) and over topography ($\alpha_2$), the scale-selective diffusion ($\tau_h^{-1}$), temperature relaxation ($\tau_r$), and the topographic scale height ($h_0^{-1} $).

The summarized derivation of the model previously included in \cite{wiegerinck_2017} and \cite{schevenhoven2017efficient} can also be found in \cref{sec:qg}.

\subsection{Training and Climate Simulation}

One instantiation of the atmospheric model with parameters $\p^0$ represents the ``truth" and was run for $11\times5000=55000$ days. In order to demonstrate the robustness of the methods used in this paper, we apply noise to this data of to represent observational errors. The noise is a simple uniform $10\%$ smearing. When nudging and training, the models are only ever using data to which this noise has been applied, which will call the ``observations".

\begin{figure*}
    \centering
    \includegraphics[width=0.59\linewidth,trim={0 0 0 0em},clip]{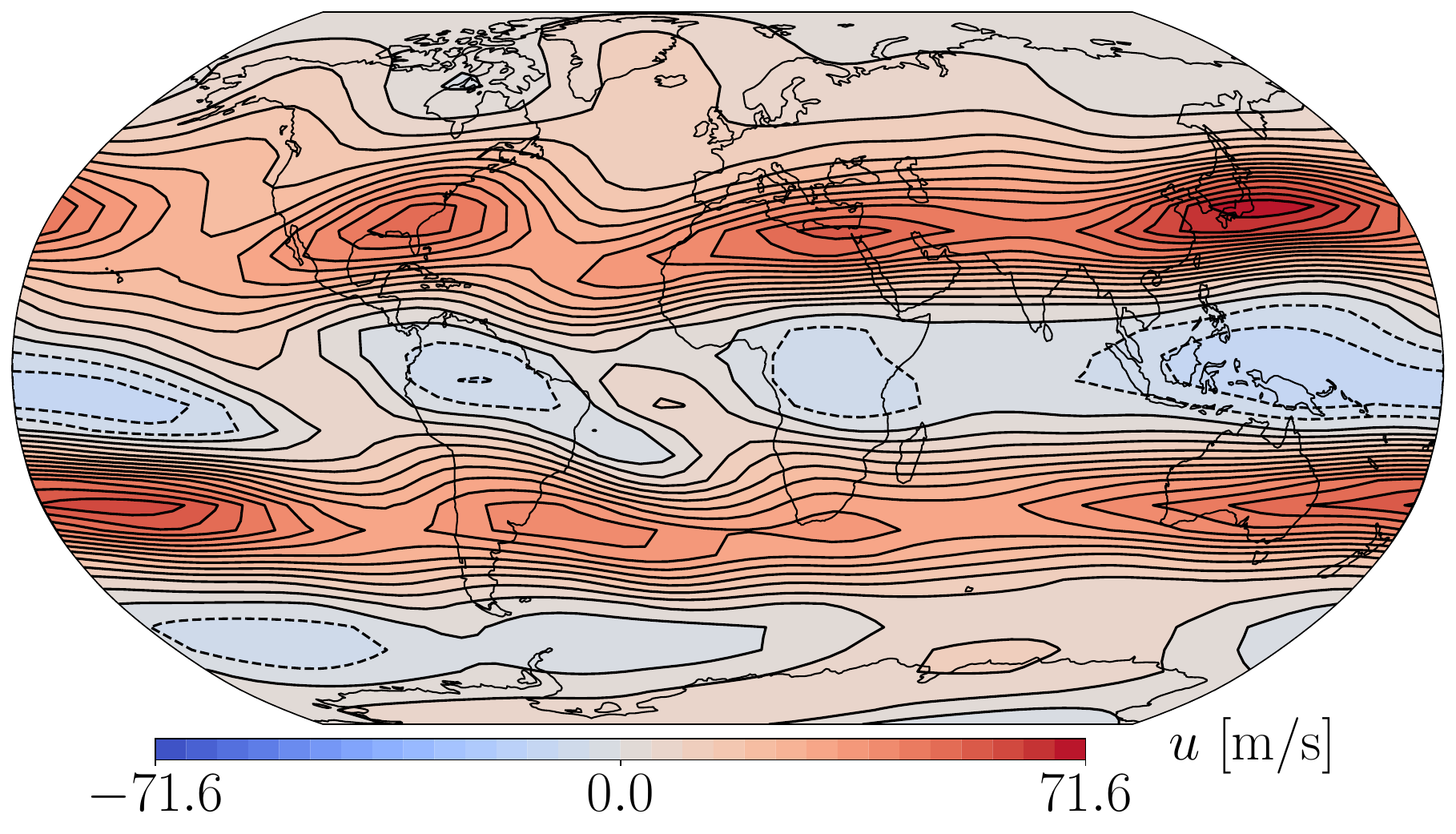}
    \includegraphics[width=0.59\linewidth,trim={0 0 0 0em},clip]{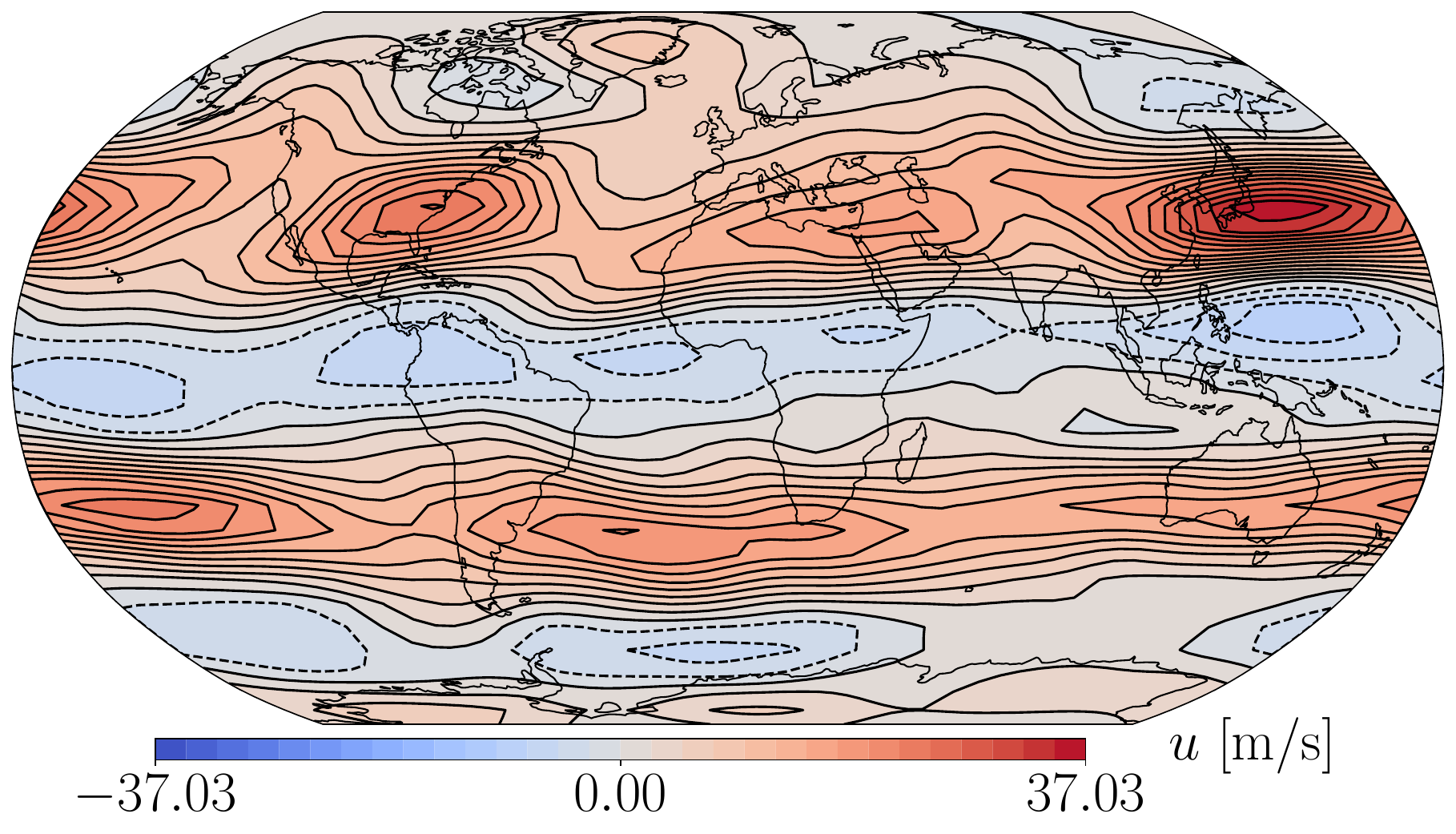}
    \includegraphics[width=0.59\linewidth,trim={0 0 0 0em},clip]{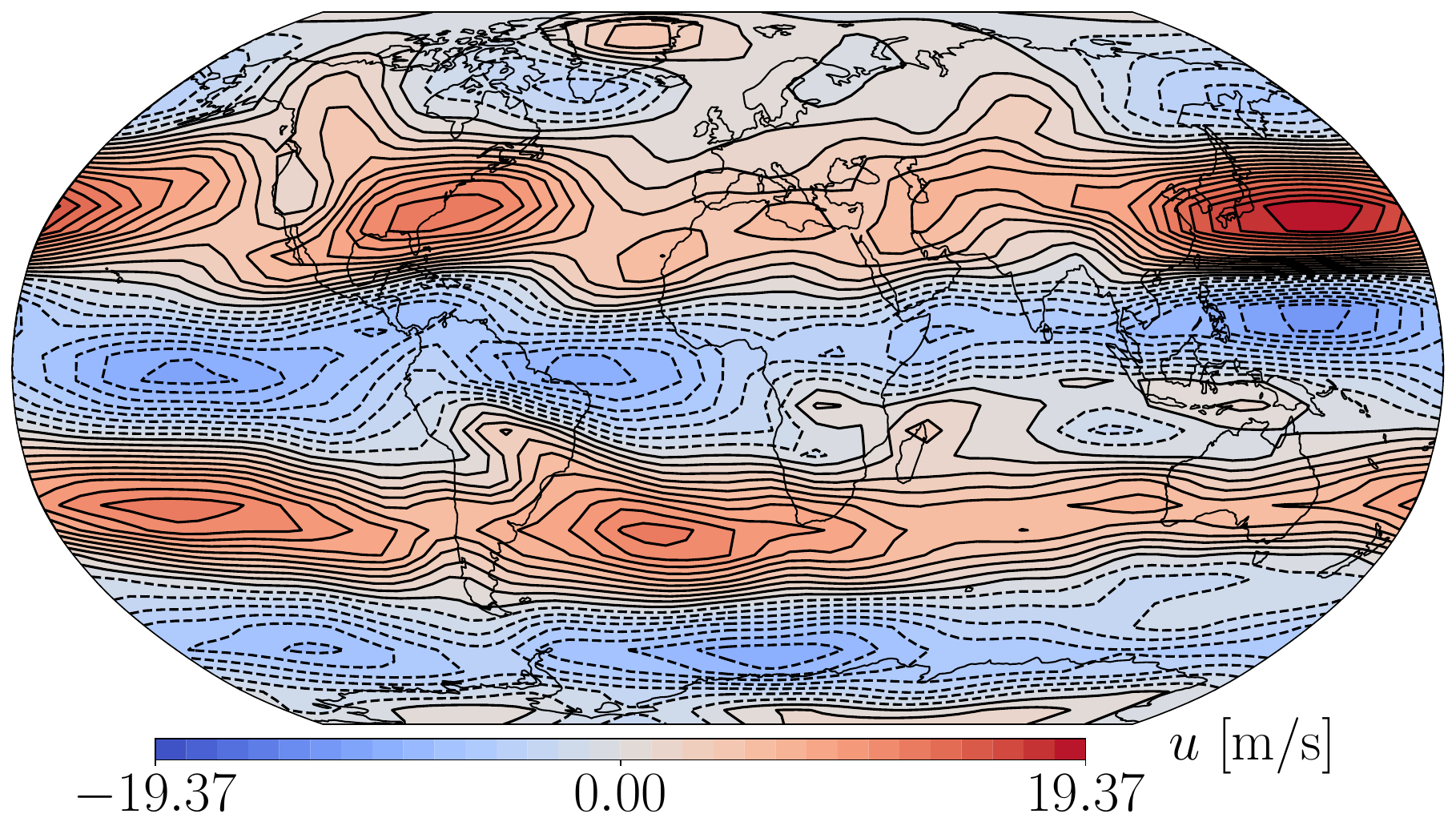}
    \caption{Mean zonal winds of QG model of ten 5000 day runs at 200 hPa, 500 hPa, and 800 hPa, going down. }
    \label{fig:clim_model}
\end{figure*}

In each experiment, the models are first run freely from perturbed initial states for 100 days to dissipate dependence on the initial conditions. They are then run in a ``nudged" configuration for 1500 days, where the state of the model is adjusted at each timestep towards the observations, resulting in soft synchronization to the truth system. Updating the model state on the basis of the observations mimics the process of data assimilation which takes place in operational weather forecasting. During the nudging period, training is applied after 100 days for the remaining 1400 days. Nudging and training is then turned off and 10 different free runs of 5000 days are integrated, all with identical initial conditions reset to the observations. These 10 runs represent $\sim137$ years of data in total which constitutes the sample from which the model climatologies are extracted. In \cref{fig:clim_model}, for illustrative purposes, the climatologies of the zonal winds of the truth are shown for each pressure level. While the training occurs at the prognostic level, with potential vorticity, we show only the zonal wind states of the model in this paper. This choice was made to present climate states in a more intuitive way than potential vorticity. Additionally, the model is trained on all three levels, but we will from now on only show the climatologies of the 500 hPa level for brevity.

The strength of the nudging to the truth $K$ is a free parameters that is chosen through trial and error with efficiency and accuracy in consideration. It must be high enough to achieve synchronization, but weak enough as not to train on random fluctuations in the data.  A nudging timescale of 60 hours is chosen.

\section{ Optimization of Model Parameters }\label{sec:opt_parameters}

We start with the optimization of 6 parameters of a model with starting values with large error. In this idealized case, the algorithm has the opportunity to find the true set of parameter values.

\subsection{Update Rules of Model Parameters}

We start by choosing the Lyapunov function $L_0\equiv \left(\Q-\Q^0\right)^2$. The rule of \cref{eq:osperule} then reduces to 
\begin{equation}
\label{eq:qdotrule}
U_{p_j} =  -2 \sum_i \left(Q_i-Q^0_i\right) \frac {\partial F_i(\Q;\p)}{\partial p_j},
 \end{equation} 

The parameter update rules for $\tau_E^{-1}$, $\alpha_1$, $\alpha_2$, $\tau_h^{-1}$, $\tau_r$, and $h_0^{-1} $ are $U_p \equiv -2 \sum_{k=1}^{483} g_k$ where $k$ runs through $m$ and $n$, and where for each parameter,\\
$g_k=$
\[ \begin{cases} 
      \dot{\tau}_{E}^{-1}:& -\tau_E(Q_{3k}-Q^0_{3k}) \left<\bk \cdot \nabla \times \left[c_d(\lambda,\phi)\bv_{\psi_3}\right],Y_{k}\right>\\
\dot{\alpha_1}:& -\tau_E^{-1}(Q_{3k}-Q^0_{3k}) \left<\bk \cdot \nabla \times [M(\lambda,\phi)\bv_{\psi_3}],Y_{k}\right> \\
\dot{\alpha_2}:& -\tau_E^{-1}(Q_{3k}-Q^0_{3k}) \left<\bk \cdot \nabla \times [H_d(\lambda,\phi)\bv_{\psi_3}],Y_{k}\right> \\
\dot{\tau}_h^{-1}:& \sum_{l=1}^{3}\tau_h\,c_h[n(n+1)]^{\frac{p_h}{2}}(Q_{lk}-Q^0_{lk})Q^{'1}_{lk}\\
\dot{\tau}_r^{-1}:&R_1^{-2}(Q_{1k}-Q^0_{1k})({\psi}_{1k}-{\psi}_{2k}) \\
                 &\   -R_1^{-2}(Q_{2k}-Q^0_{2k})({\psi}_{1k}-{\psi}_{2k}) \\
                 &\   +R_2^{-2}(Q_{2k}-Q^0_{2k})({\psi}_{2k}-{\psi}_{3k}) \\
                 &\   -R_2^{-2}(Q_{3k}-Q^0_{3k})({\psi}_{2k}-{\psi}_{3k}) \\
\dot{h}_0^{-1} :&(Q_{3k}-Q^0_{3k})\left<-\bv_{\psi_3}\cdot\nabla f h,Y_{k}\right>, \label{eq:parupdate}
   \end{cases}
\]
with $\left<\mathcal{F},Y_{k}\right>$ the projection of some field $\mathcal{F}$ onto the spherical harmonical function $Y_{k}$.

\subsection{ Experiment }
\begin{figure}
    \centering
    \includegraphics[width=\linewidth]{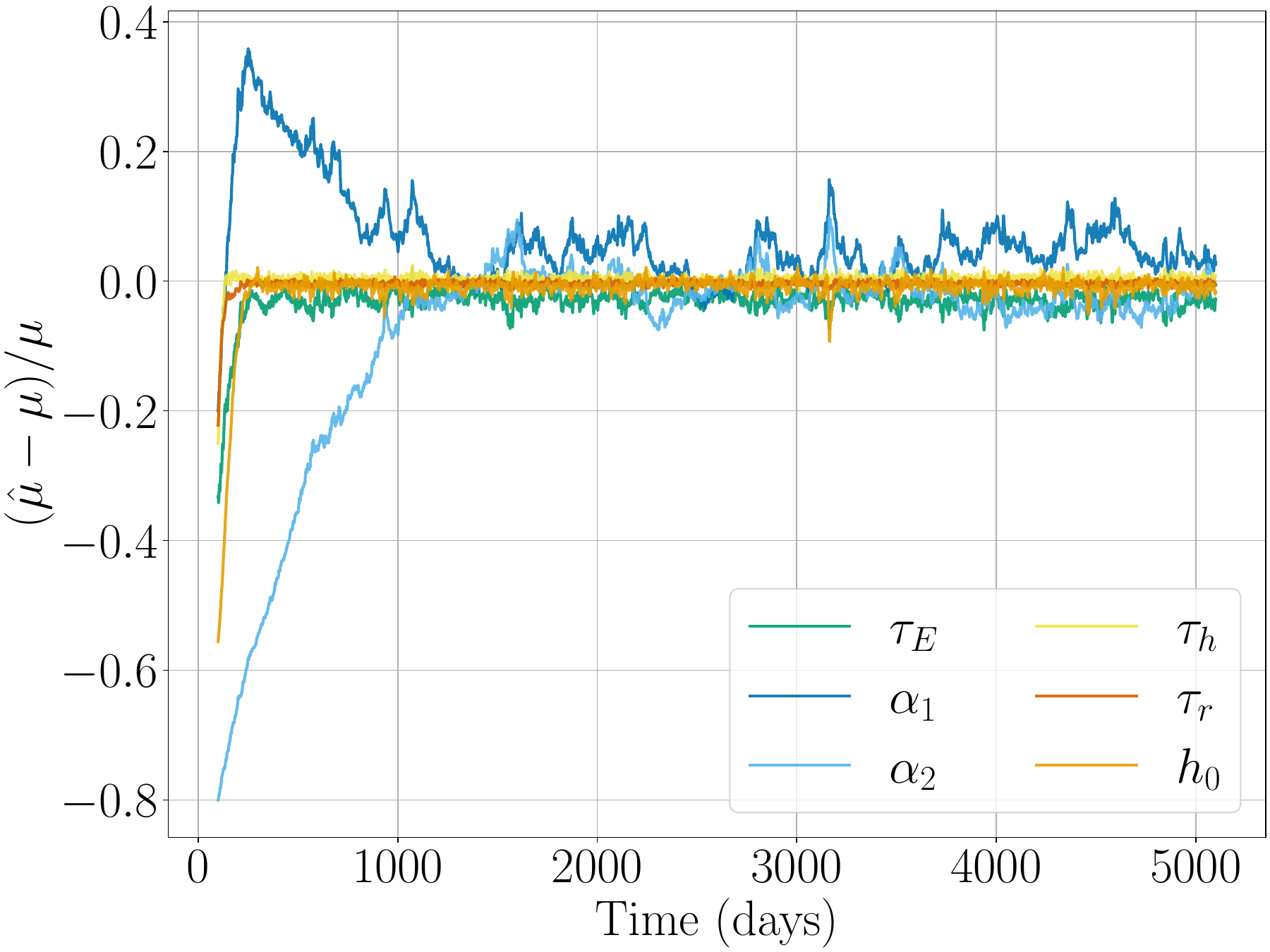}
    \caption{After 1500 days of integration, the parameters have stabilized to $\sim 20\%$ error. } 
    \label{fig:sync_params}
\end{figure}

\begin{figure}
    \centering
    \includegraphics[width=\linewidth,clip]{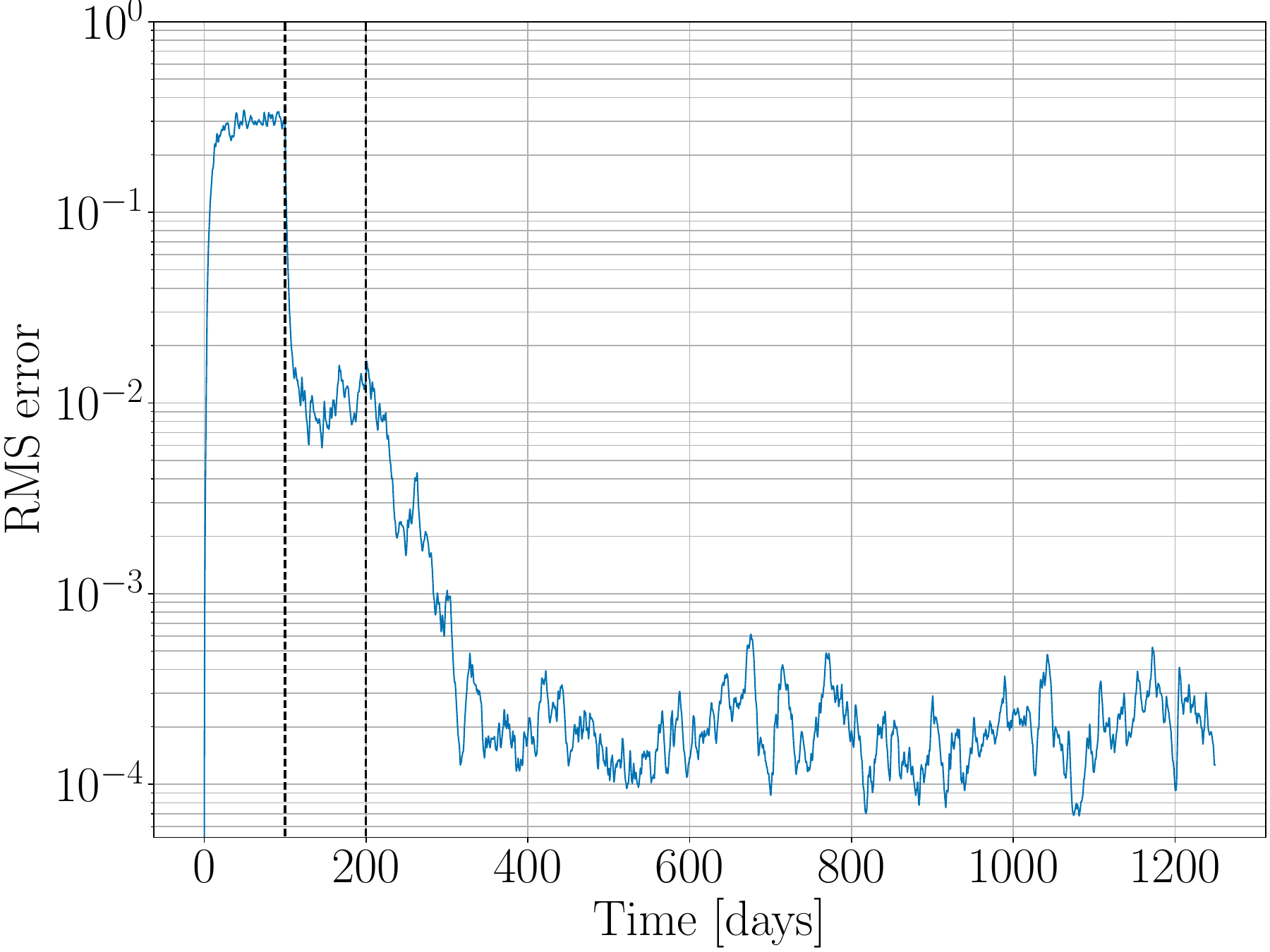}
    \caption{One timestep forecast synchronization error. The dotted lines indicate first the start of nudging, followed by the start of training. During nudging, the error lies at $~10^{-2}$, which reduces to $<3\times10^{-4}$ during training.}
    \label{fig:syncherr}
\end{figure}

We define model $\eta$ with parameters $\p^\eta$ all set to be underestimated. During training, parameters converge to the values of the truth model. A longer training of 5000 days was performed to illustrate the long term stability of the method in \cref{fig:sync_params}. 
The starting and final values of the parameters are shown in \cref{tab:parameters_30}.
\begin{table}[hbtp]
\begin{center}
\begin{tabular}{p{3 cm} |lll }
   & $\p^0$ & $\p^\eta$  & $\p^{\eta}{'}$  \\
 \hline
 \hline
$\tau_E$ [days] & 4.5 & 3.0 & $4.36 \pm 0.04$ \\
\hline
$\alpha_1$ & 0.5 & 0.4 & $0.51 \pm 0.02$ \\
\hline
$\alpha_2$ & 0.5 & 0.1 & $0.52 \pm 0.03$  \\
 \hline
$\tau_h$ [days] & 4 & 3.0 &  $4.049 \pm 0.019$  \\
 \hline
$\tau_r$ [days] & 45 & 35 & $44.88 \pm 0.14$ \\
\hline
 $h_0 $ [km] & 9.0 & 4.0 &  $8.88 \pm 0.07$ \\
\end{tabular}
\caption{ Parameter values of the truth, imperfect, and trained models. The errors indicate the standard deviation of the last half values in the training time sequence.}
\label{tab:parameters_30}
\end{center}
\end{table}
As the parameter error decreases, the synchronization error of 1 timestep forecasts also decreases, see \cref{fig:syncherr}.

\begin{figure}
    \centering
    \includegraphics[width=\linewidth,trim={0 0 0 0em},clip]{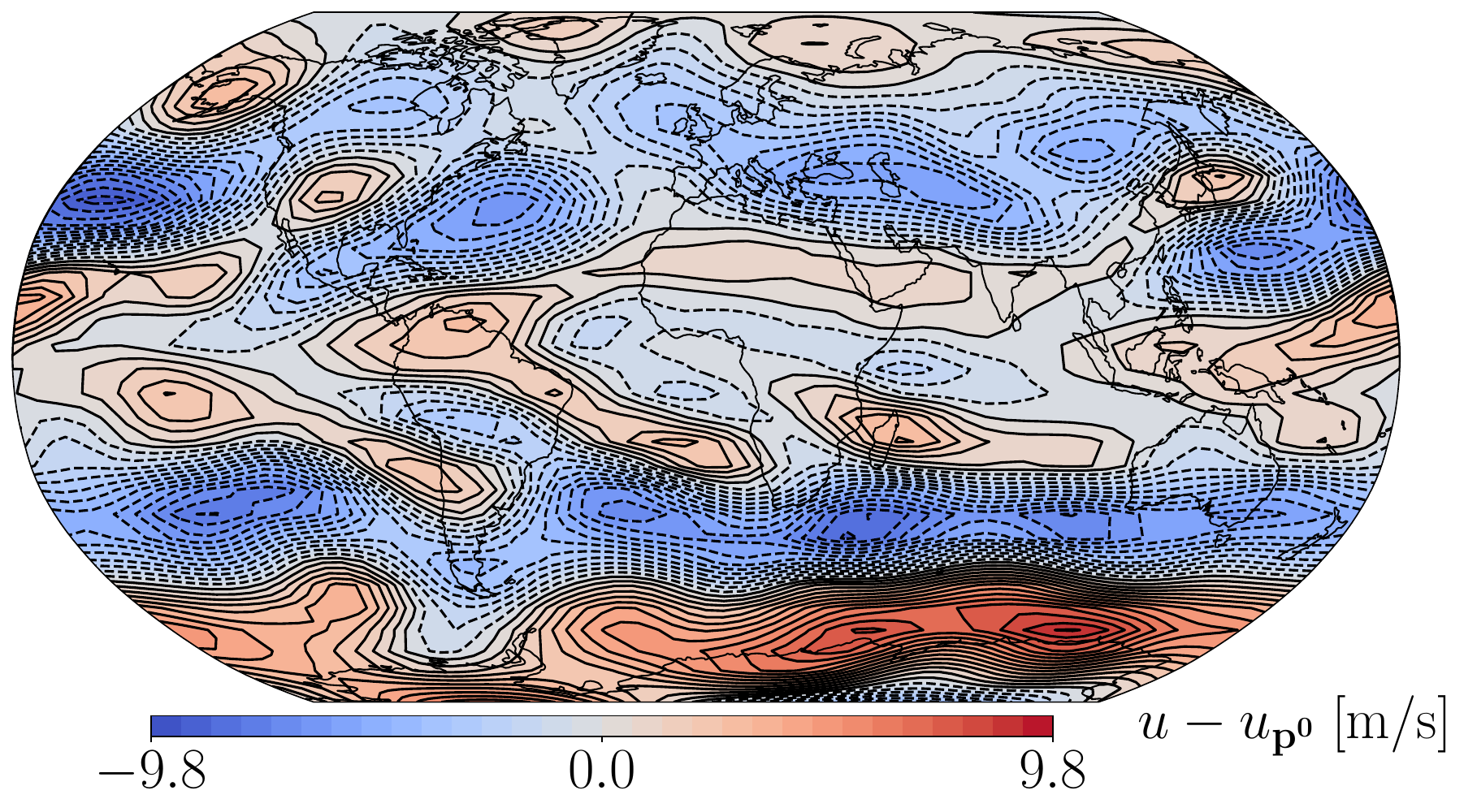}
    \includegraphics[width=\linewidth]{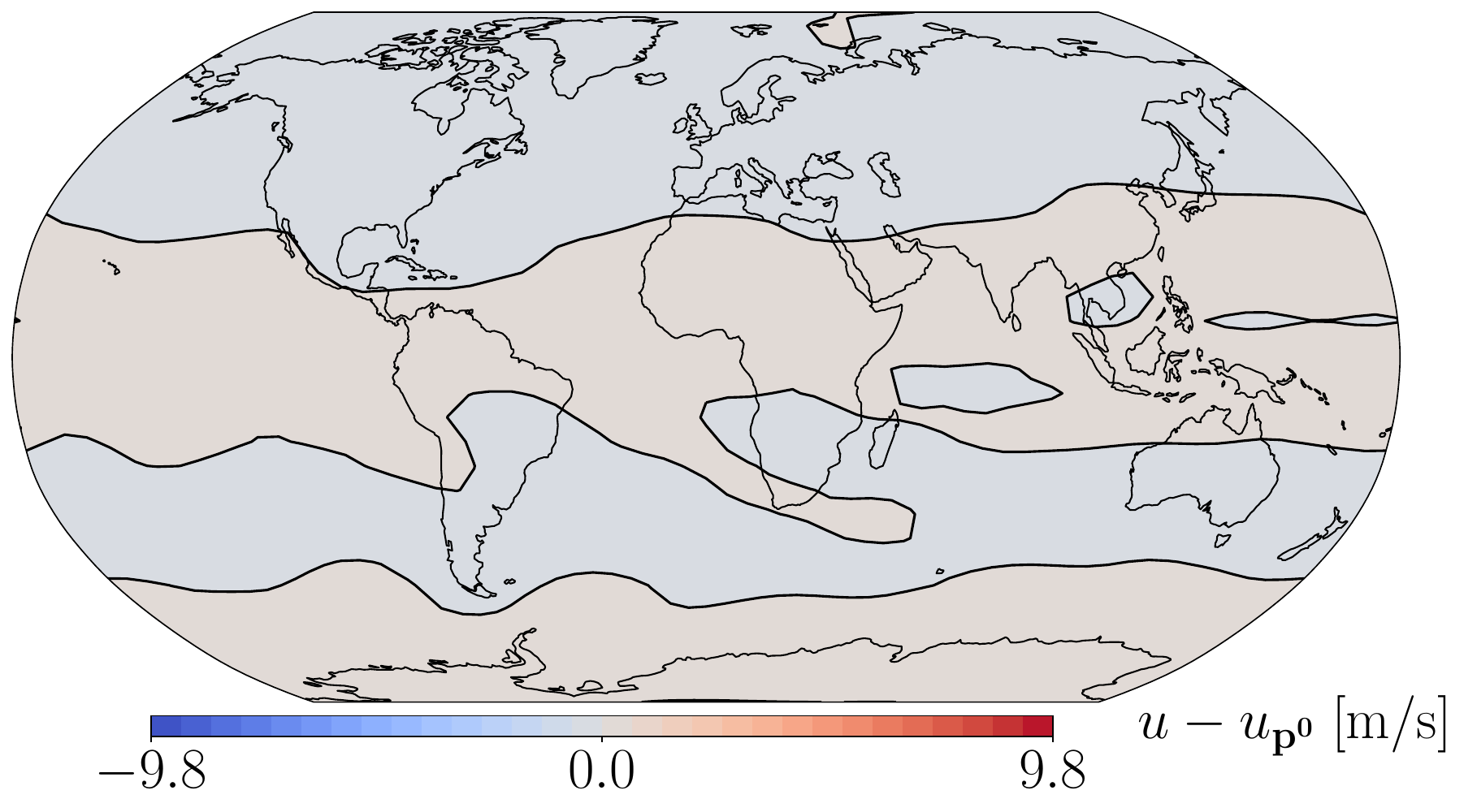}
    \caption{Difference between truth and model climatology at 500 hPa before (top) and after (bottom) training. The error in the climatology of the trained model does not exceed the 0.5 m/s level.}
    \label{fig:param_clim_diff}
\end{figure}

The improvements in single timestep forecasts translate to the long term forecasting skill. The errors of the climatological zonal winds at the 500 hPa level reduce from $O(10)$ m/s at worst to $~0.5$ m/s or less globally, see \cref{fig:param_clim_diff}. 
The spatial RMSE of the temporal mean and standard deviation of the zonal winds at the 500 hPa level are reported in \cref{tab:rmse_30}. Note that the RMSE of perfect model 0 does not vanish when starting from different initial conditions. This is due to an orbit which is either open or has a period exceeding the integration time due to the large dimensionality of the system. The performance of the imperfect models is instead compared to that of the model 0, which is taken to be the best possible performance.
\begin{table}[]
\begin{tabular}{lll}
\toprule
 & RMSE $\left<u\right>$ & RMSE $\sigma u$ \\
\midrule
Model 0 & $0.53 \pm 0.04$ & $0.219 \pm 0.006$ \\
Model $\eta$ & $3.12 \pm 0.02$ & $1.569 \pm 0.016$ \\
Model $\eta'$ & $0.52 \pm 0.03$ & $0.226 \pm 0.012$ \\
\bottomrule
\end{tabular}
\caption{ Root mean squared error of the temporal mean and standard deviation of zonal winds of all cells at the 500 hPa level with respect to the truth. }
\label{tab:rmse_30}
\end{table}

\section{ Optimization of Supermodel Weight }\label{sec:sumo}

Errors in parameter values are not the only sources of model bias. Structural errors in the approximate representation of physical processes or missing physics will also reduce model performance. These can be reduced by the use of a \emph{supermodel} (SUMO). 
We now present a case of two member models with parameter errors, but where the training of parameters is disallowed to represent structural errors which cannot be corrected by parameter tuning. We overcome this by the use of a weighted SUMO. This experiment is also idealized, as we set large differences between the structural error biases of each model. If the member models were identically biased, adjusting the weighting of the SUMO would have no effect. We now optimize the SUMO member weights rather than the model parameters.


\subsection{Update Rule of 2-member SUMO Weight}

\begin{figure}
    \centering
    \includegraphics[width=0.5\linewidth]{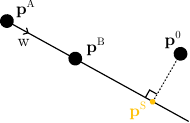}
    \caption{Shortest distance between $\p^0$ and $\p^S$ by setting $w$ according to \cref{eq:w_opt}.}
    \label{fig:sumo_parameters}
\end{figure}

Take some models, $A$ and $B$, with parameters $\p^A$ and $\p^B$. The tendency of the SUMO is defined as
\begin{equation}\label{wsumo_def}
    F^S_i(\Q^S,w) \equiv (1-w) F_i(\Q^S, \p^A) + wF_i(\Q^S, \p^B),
\end{equation}
where $w$ is a weight which can be optimized, $F_i^S$ is the tendency at index $i$, and $\Q^S$ is the state of the SUMO. 
To first degree\footnote{See Taylor's theorem for multivariate functions.},
\begin{align} 
    F_i(\Q^S,\p^A+\boldsymbol{\Delta}) \simeq F_i(\Q^S,\p^A) + \sum_j^n\frac{\partial F_i(\Q^S,\p^A)}{\partial p_j}\Delta_j,
\end{align}
where $\boldsymbol{\Delta} \equiv \p^B - \p^A$ and $n$ is the number of parameters. 
We can now rewrite \cref{wsumo_def} as
\begin{align}\label{eq:eff_params}
    F^S_i(\Q^S,w) \simeq & F_i(\Q^S,\p^A) + w\sum_i^n\frac{\partial F_i(\Q^S,\p^A)}{\partial p_i}\Delta_i\notag\\
    \simeq & F_i\left(\Q^S,\p^S(w)\right)
\end{align}
where $\p^S(w) \equiv (1-w)\p^A + w\p^B$ is the vector of \emph{effective} SUMO parameters. 
If there exists a weight $w^t$ such that $\p^S(w^t) = \p^0$, the conditions for identical synchronization are fulfilled. $L_0(w^t) \equiv \left(\Q^S - \Q^0\right)^2$ can then be chosen as a Lyapunov function.
The derivative of the tendency error with respect to the weight is
\begin{align}
     \frac{\partial h_i}{\partial w} & = F_i(\Q^S, \p^B) - F_i(\Q^S, \p^A),
\end{align}
and the update rule is
\begin{align}\label{weight_rule}
U_w &=  -2\sum_i \left(Q^S_i-Q^0_i\right) \left(F_i(\Q, \p^B) - F_i(\Q, \p^A) \right).
\end{align}
Note here that the weight need not remain within 0 and 1. A value $w\notin[0,1]$ lets the SUMO adopts effective parameters which are larger (or smaller) than both models \cite{negative_weights_schevenhoven}. Further, if no $w^t$ exist such that $\p^S(w^t)=\p^0$, one could expect $w$ to instead get $\p^S$ as close to $\p^0$ as possible. This is achieved by setting
\begin{equation}
    w=\frac{(\p^B-\p^A)(\p^0-\p^A)}{|\p^B-\p^A|^2},
    \label{eq:w_opt}
\end{equation}as represented in \cref{fig:sumo_parameters}.
Two cases where $w\in[0,1]$ and $w\notin[0,1]$ are explored in the next section.

\subsection{ Experiment }\label{sec:exp_sumo}
Three models, 1, 2 and 3, are defined, the parameters of which differing from $\boldsymbol{p}^0$ are shown in \cref{tab:parameters_sumo}. Two experiments are set up, one with SUMO$_{12}$ (SUMO of model 1 and 2), and SUMO$_{13}$. 

\begin{table}[hbtp]
\begin{center}
\begin{tabular}{p{3 cm} |lllll }
   & $\p^0$ & $\p^1$  & $\p^2$ & $\p^3$  \\
 \hline
 \hline
$\tau_h$ [days] & 4.0 & 7.0 & 3.0 & 9.0 \\
 \hline
$\tau_r$ [days] & 45 & 20 & 100 & 10 \\
\end{tabular}
\caption{ Parameter values of the truth and imperfect models, where they differ.}
\label{tab:parameters_sumo}
\end{center}
\end{table}

\begin{figure}
    \centering
    \includegraphics[width=\linewidth,clip]{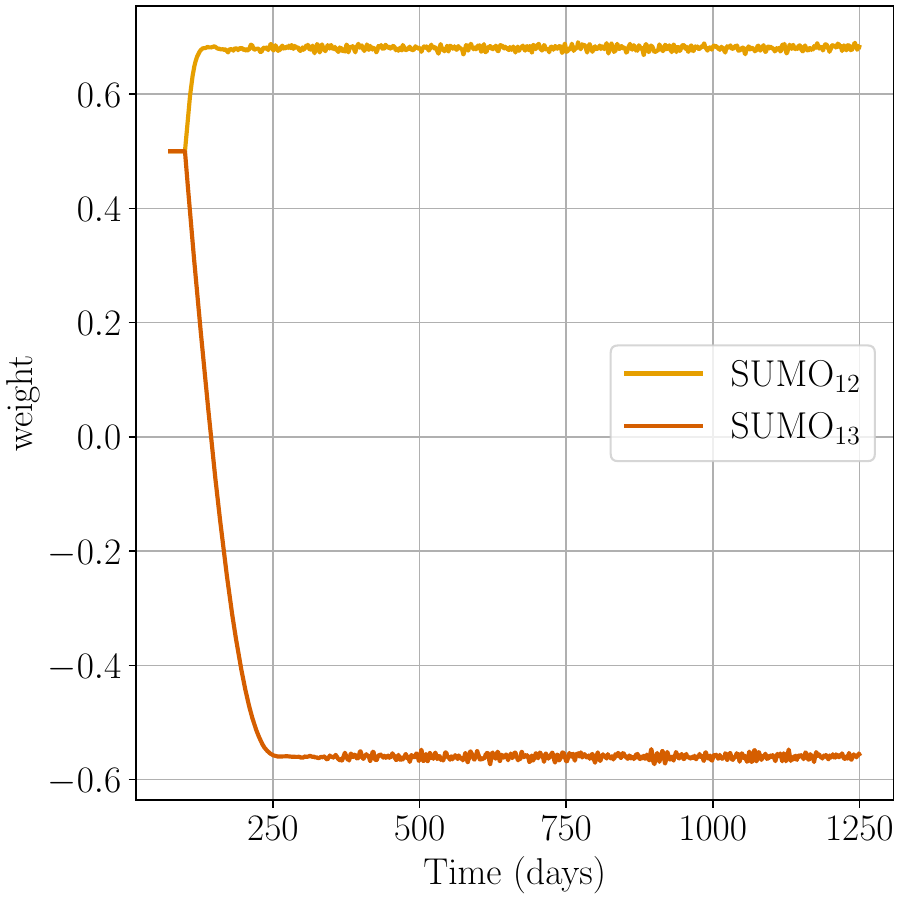}
    \caption{SUMO weight with training started at $t=100$ days.}
    \label{fig:train_weight}
\end{figure}

The weights stabilize within 250 days of training, see \cref{fig:train_weight}. Since $p^0_j$ lies between $p^1_j$ and $p^2_j$ for every parameter $j$, the weight of SUMO$_{12}$ optimizes to $w\in[0,1]$ to find effective parameter values close to that of $p^0_j$, as expected from \cref{eq:w_opt}. In the case of SUMO$_{13}$, $p^0_j$ does not lie between $p^1_j$ and $p^2_j$, and a solution where $w < 0$ is found. In both cases, the effective parameter values of the SUMO are closer to those of the truth model than any of the member models. The final values of the weights and effective parameter values are reported in \cref{tab:parameters_sumo_eff}. 
\begin{table}[hbtp]
\begin{center}
\begin{tabular}{l |ll }
   & SUMO$_{12}$ & SUMO$_{13}$  \\
 \hline
 \hline
$\tau^S_h$ [days] & $4.272 \pm 0.014$ & $5.876 \pm 0.010$  \\
 \hline
$\tau^S_r$ [days] & $74.6 \pm 0.3$ & $25.62 \pm 0.04$  \\
\hline
$w$ & $0.682\pm0.004$ & $-0.562 \pm 0.004$
\end{tabular}
\caption{ Effective parameter values of the SUMOs and their weights. The errors indicate the standard deviation of the last half values in the training time sequence.}
\label{tab:parameters_sumo_eff}
\end{center}
\end{table}

\begin{figure}
    \centering
    \includegraphics[width=\linewidth]{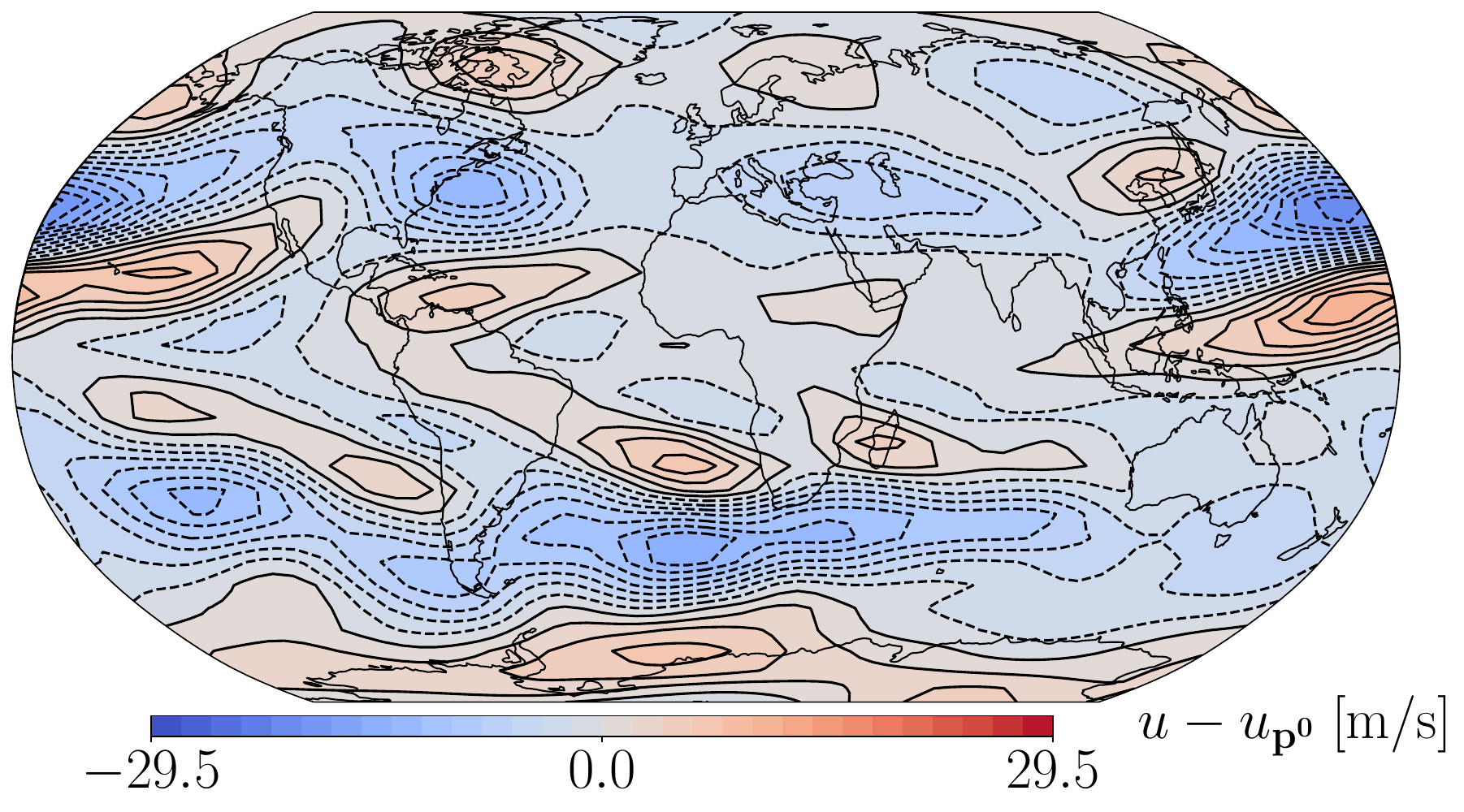}
    \includegraphics[width=\linewidth]{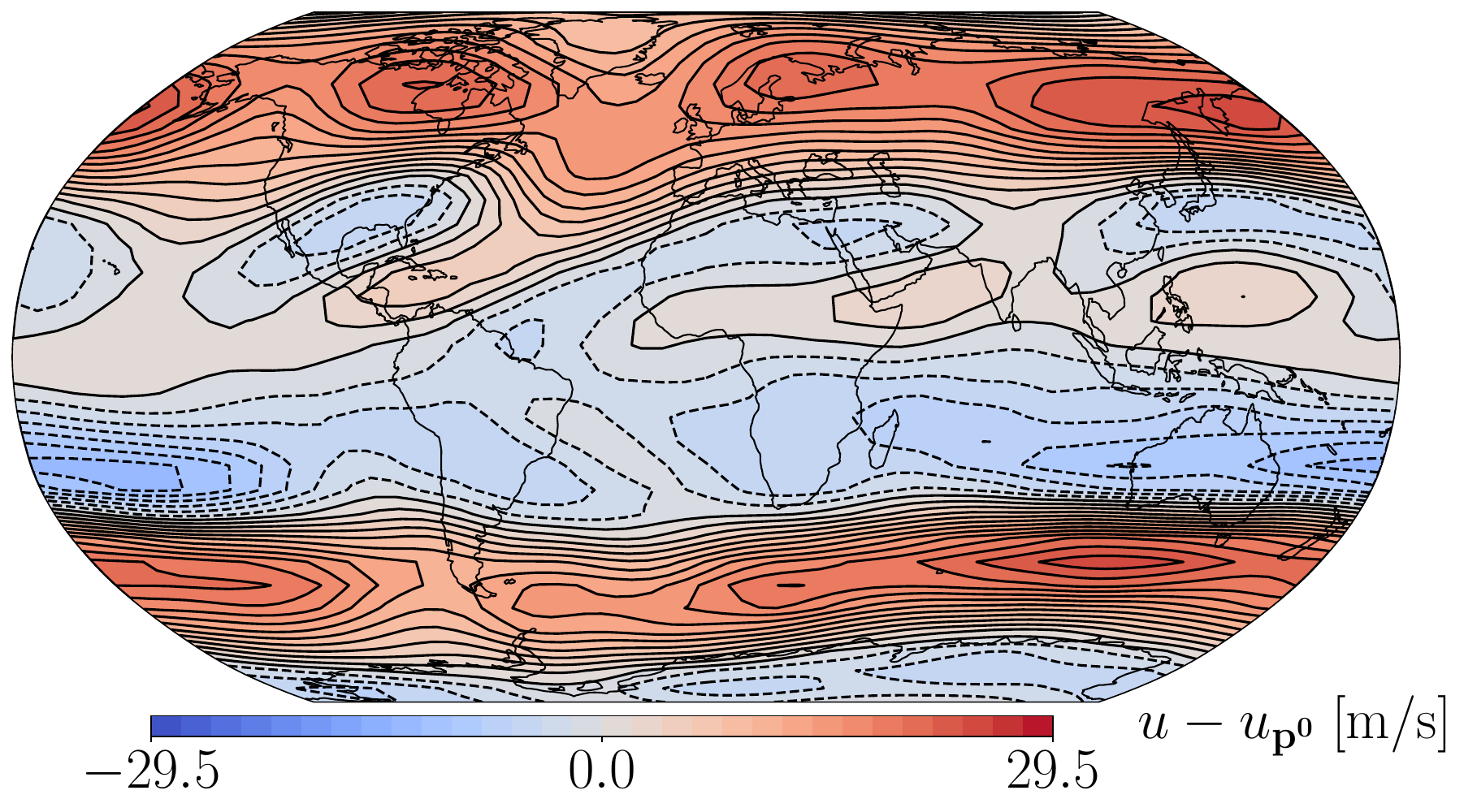}
    \includegraphics[width=\linewidth]{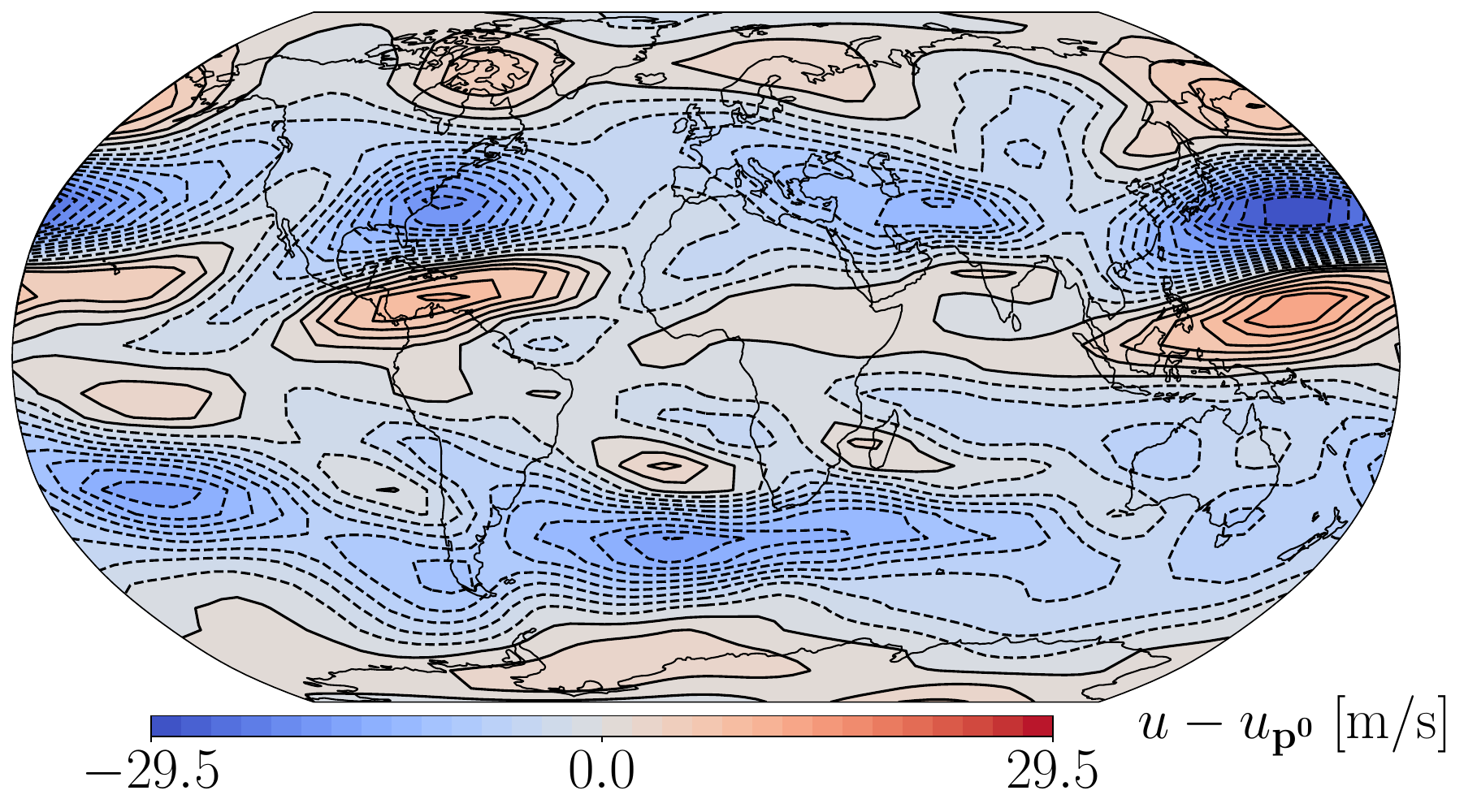}
    \caption{Error of climatological zonal winds at the 500 hPa level of model 1, 2, and 3, going down.}
    \label{fig:sumo_member_clim_diff}
\end{figure}

\begin{figure}
    \centering
    \includegraphics[width=\linewidth]{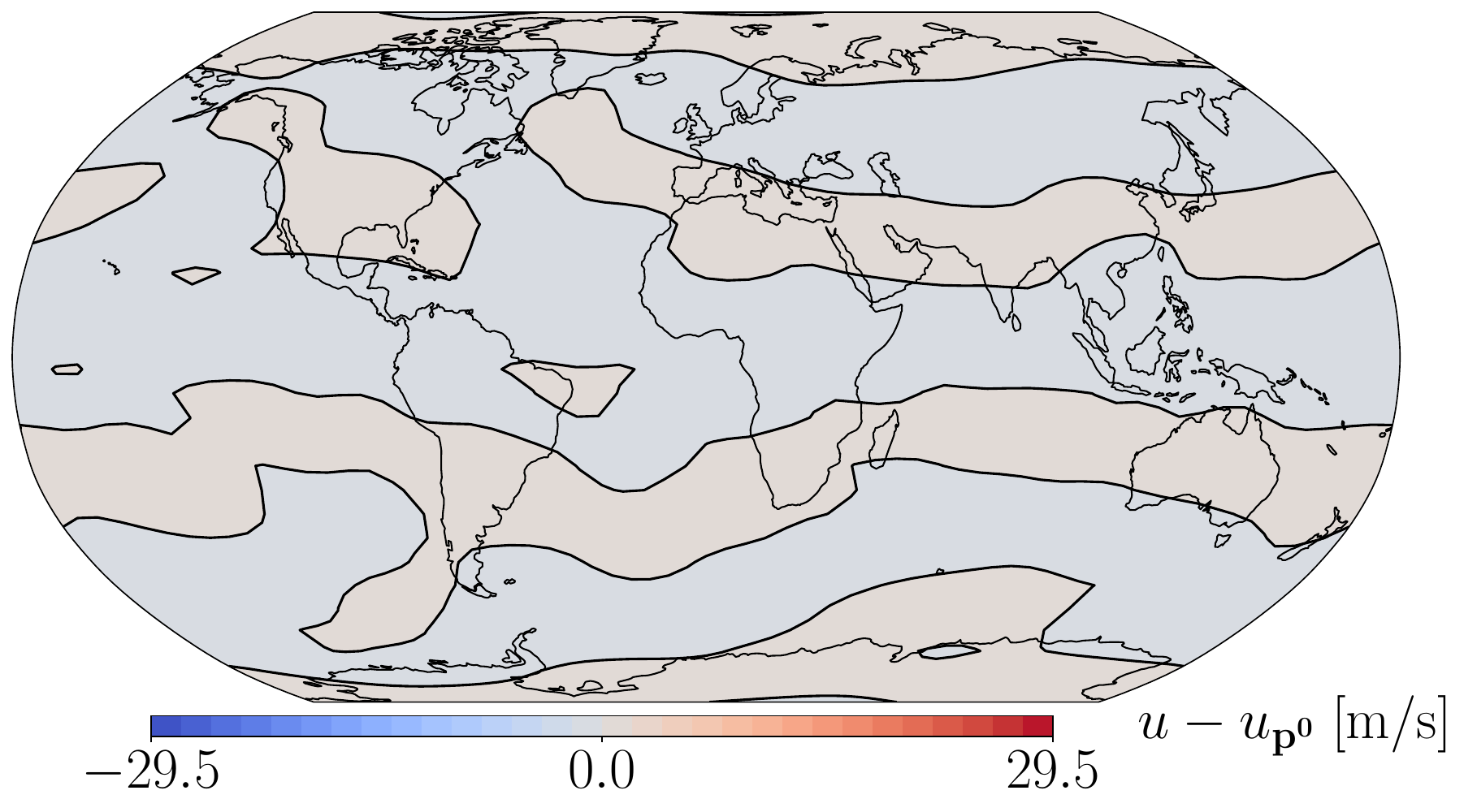}
    \includegraphics[width=\linewidth]{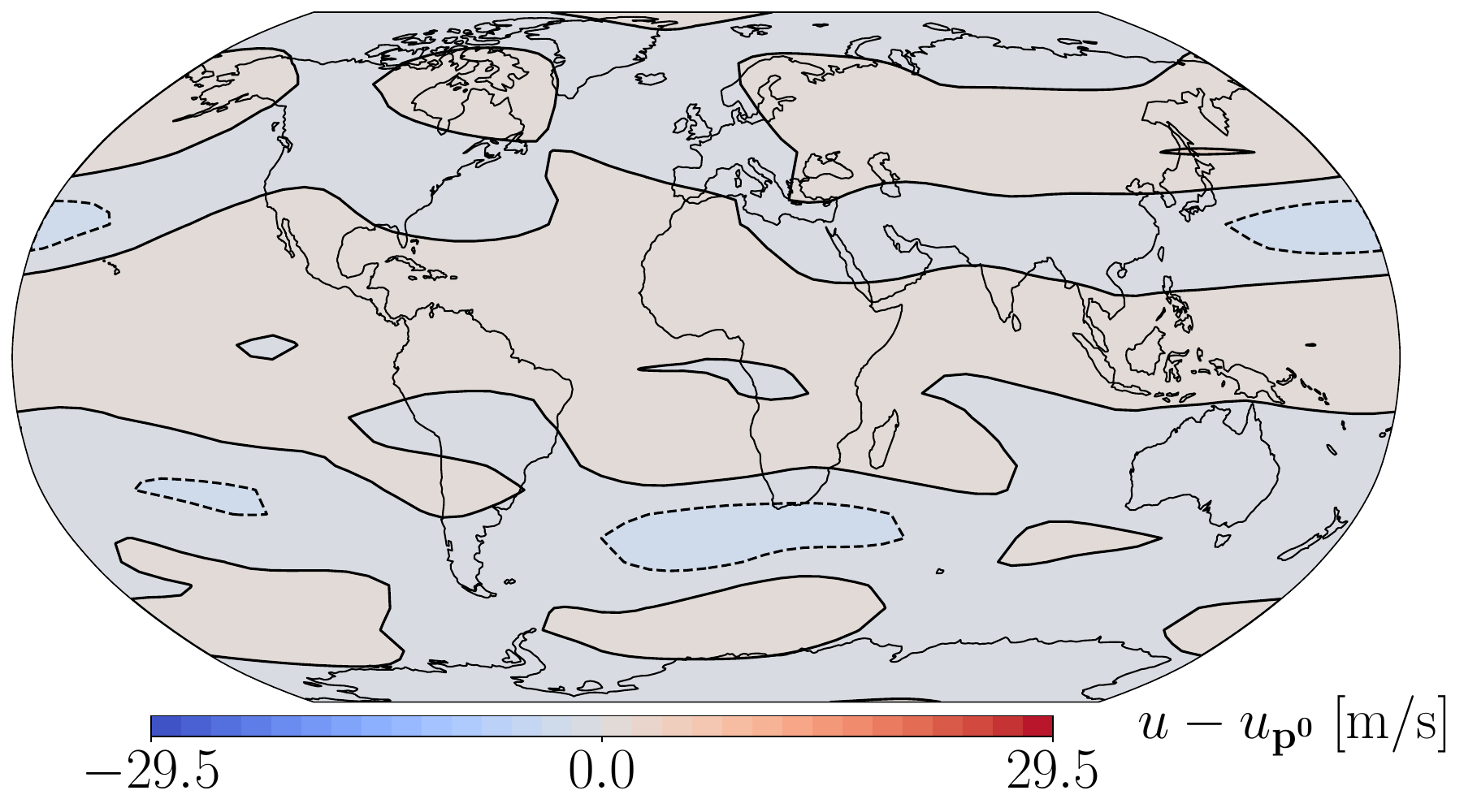}
    \caption{Error of climatological zonal winds at the 500 hPa level of SUMO$_{12}$ and SUMO$_{13}$, going down.}
    \label{fig:sumo_clim_diff}
\end{figure}

Despite model $2$ and $3$ being overall worse than model 1, the SUMOs which include them yield a globally reduced error in the climatological zonal winds (\cref{fig:sumo_clim_diff}) compared to any of their member models (\cref{fig:sumo_member_clim_diff}). The mean global RMSE of the temporal mean and temporal standard deviation of the zonal winds are reported in \cref{tab:rmse_sumo}.

\begin{table}
\begin{tabular}{l|ll}
\toprule
 & RMSE $\left<u\right>$ & RMSE $\sigma u$  \\
\midrule
Model 0 & $0.53 \pm 0.04$ & $0.219 \pm 0.006$ \\

Model 1 & $4.526 \pm 0.017$ & $1.629 \pm 0.015$ \\
Model 2 & $10.15 \pm 0.07$ & $ 3.58 \pm 0.02$ \\
Model 3 & $6.105 \pm 0.016$ & $2.886 \pm 0.014$ \\
SUMO$_{12}$ & $0.68 \pm 0.11$ & $0.33 \pm 0.01$ \\
SUMO$_{13}$ & $0.99 \pm 0.04$ & $0.901 \pm 0.013$ \\
\bottomrule
\end{tabular}
\caption{ Root mean squared error of the temporal mean and standard deviation of zonal winds of all cells at the 500 hPa level with respect to the truth. }
\label{tab:rmse_sumo}
\end{table}


\section{ Combined Optimization of Weight and Parameters} \label{sec:asumo}

Optimization of parameters and SUMO weights each have their weaknesses. 
Here we design an experiment which contains hurdles for both methods: non-trainable parameters (representing structural errors) and identically biased member parameters. To overcome these, we implement an ASUMO, where both the weights of the SUMO and the parameters of one of its member models are trained simultaneously. 

\subsection{Update Rules for Parameters and Weights}
We follow the same reasoning as in \cref{sec:sumo}, but we now want to modify both the weight and model parameters. Assuming there exists some combination of $w$, $\p^A$ and $\p^B$ such that $w\p^B + (1-w)\p^A = \p^0$, $L_0(w^t) \equiv \left(\Q^S - \Q^0\right)^2$, can again be chosen as a Lyapunov function.
The partial derivatives of the error in tendency are now
\begin{align}
     \frac{\partial h_i}{\partial w} & = F_i(\Q^S, \p^A) - F_i(\Q^S, \p^B) \\
     \frac{\partial h_i}{\partial p^A_j} & = (1-w)\frac{\partial F_i(\Q^S, \p^A_j) }{\partial p^A_j} \\
     \frac{\partial h_i}{\partial p^B_j} & = w\frac{\partial F_i(\Q^S, \p^B_j) }{\partial p^B_j}
\end{align} where $p^x_j$ is some parameter $p_j$ of model $x$. The update rules for the weight and model parameters are then
\begin{align}
    U_w &= -2 \sum_i \left(Q^S_i-Q^0_i\right) \left(F_i(\Q^S, \p^A) - F_i(\Q^S, \p^B) \right)\\
    U^A_{p_j} &= -2 (1-w)\sum_i  \left(Q^S_i-Q^0_i\right)\frac{\partial F_i(\Q^S, \p^A) }{\partial p^A_j}\\
    U^B_{p_j} &= -2 w\sum_i \left(Q^S_i-Q^0_i\right)\frac{\partial F_i(\Q^S, \p^B) }{\partial p^B_j},
\end{align}
where $r_w$ and $r_j$ are the learning rates of $w$ and parameter $p_j$, respectively.  

\subsection{Experiment}\label{sec:setup_asumo}
We use model $1$ from in \cref{sec:exp_sumo}, and define a model $4$ with parameters as indicated in \cref{tab:parameters_asumo}. Note that $\p^1$ and $\p^4$ are both given an underestimated and identical $\tau_r$, making it impossible to correct this bias with SUMO only. The values of $\tau_h^{-1}$ are also made non-trainable, representing structural errors which cannot be solved by parameter optimization.

The models are then integrated for 6 different experiments. Using the perfect model 0, model $1$, model $4$, model $1$ with training of parameters, SUMO$_{14}$ ( of models $1$ and $4$ ) without, and with training the parameters of model $1$, ASUMO$_{14}$.

\begin{table}[hbtp]
\begin{center}
\begin{tabular}{l|llll }
   & $\p^0$ & $\p^1$ & $\p^4$ & $\p^{1,AS}$ \\
 \hline
 \hline
$\tau_E$ [days] & 4.5 & 4.5 & 4.5 & 4.5 \\
\hline
$\alpha_1$ & 0.5 & 0.5 & 0.5 & $0.51 \pm 0.007$ \\
\hline
$\alpha_2$ & 0.5 & 0.5 & 0.5 & $0.520 \pm 0.005$\\
 \hline
$\tau_h$ [days] & 4 & 7 & 9 & 7\\
 \hline
$\tau_r$ [days] & 45 & 20 & 20 & $22.988\pm 0.011$\\
\hline
 $h_0 $ [km] & 9 & 9 & 9 & $8.97\pm0.03$\\
\end{tabular}
\caption{ Parameter values of the perfect, imperfect, and trained models. The errors indicate the standard deviation of the last half values in the training time sequence.}
\label{tab:parameters_asumo}
\end{center}
\end{table}

When training both parameters and SUMO weights, synchronization occurs after a longer period of time for the weights, 800 days compared to and 250 days in \cref{sec:sumo}. The parameter values of model 1 as a member of ASUMO$_{14}$ after training are shown in \cref{tab:parameters_asumo}. ASUMO$_{14}$ finds an optimal weight $w=-3.370\pm0.002$.
\begin{figure*}
\begin{subfigure}[t]{0.5\linewidth}
    \includegraphics[width=\linewidth]{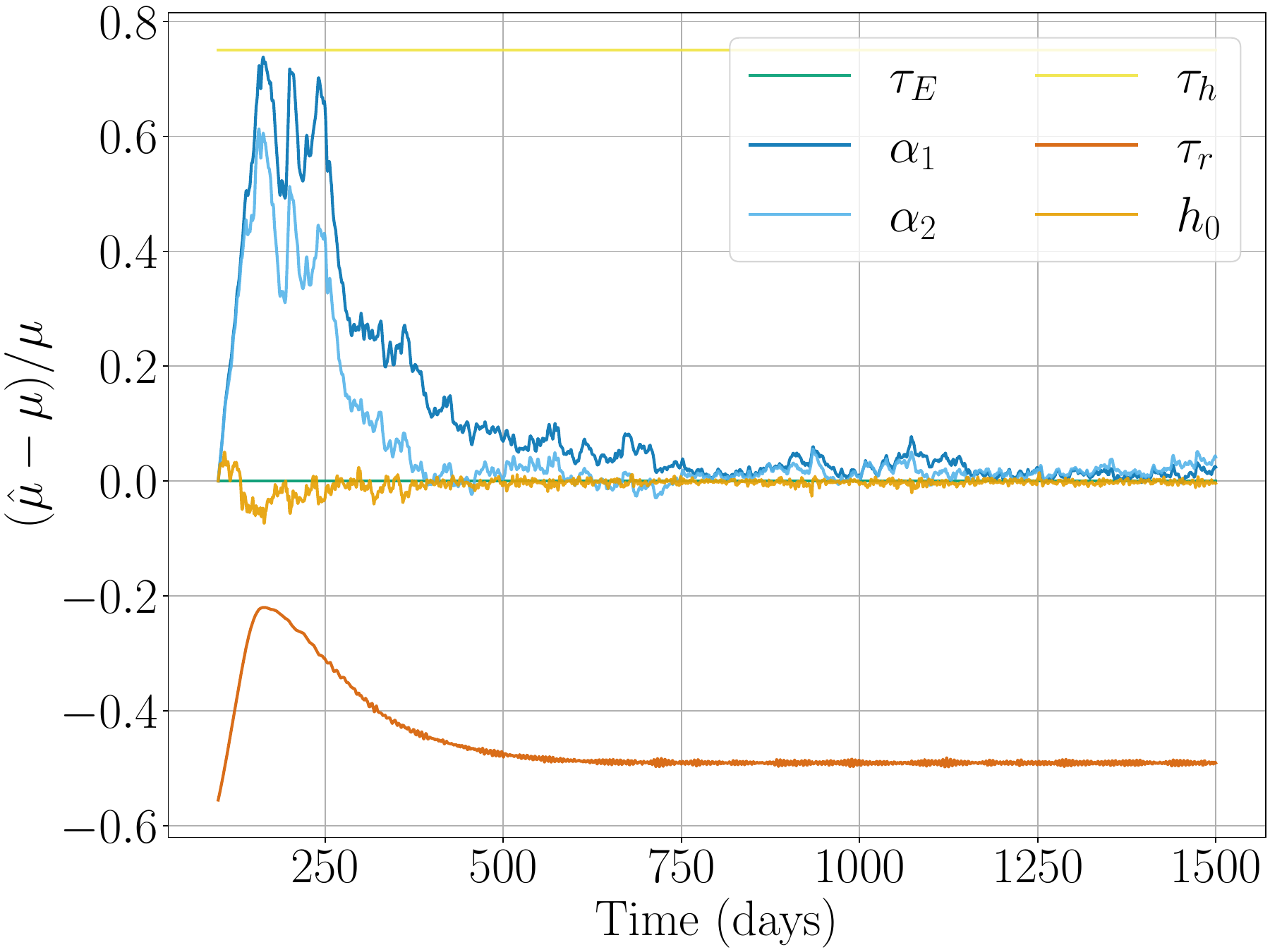}
    \caption{ Parameter errors of model 1 as a member of ASUMO$_{14}$. }
\end{subfigure}
~
\begin{subfigure}[t]{0.5\linewidth}
    \includegraphics[width=\linewidth]{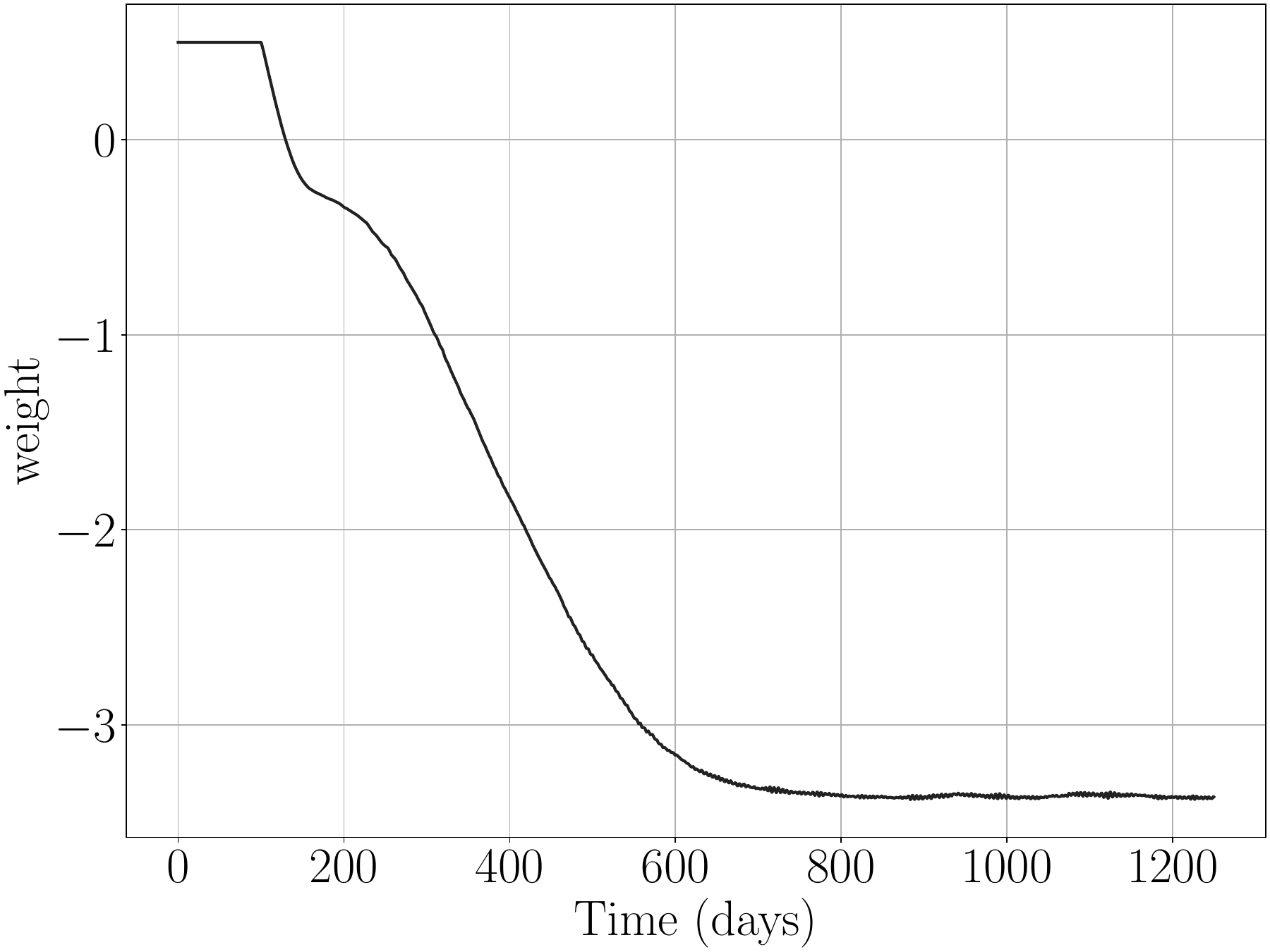}
    \caption{ Weight.}
\end{subfigure}    
\caption{ Training of parameters of ASUMO$_{14}$.}
\label{fig:params_train_asumo}
\end{figure*}





\begin{figure*}[t!]
    \centering
    \begin{subfigure}[t]{0.50\linewidth}
        \centering
        \includegraphics[width=\linewidth]{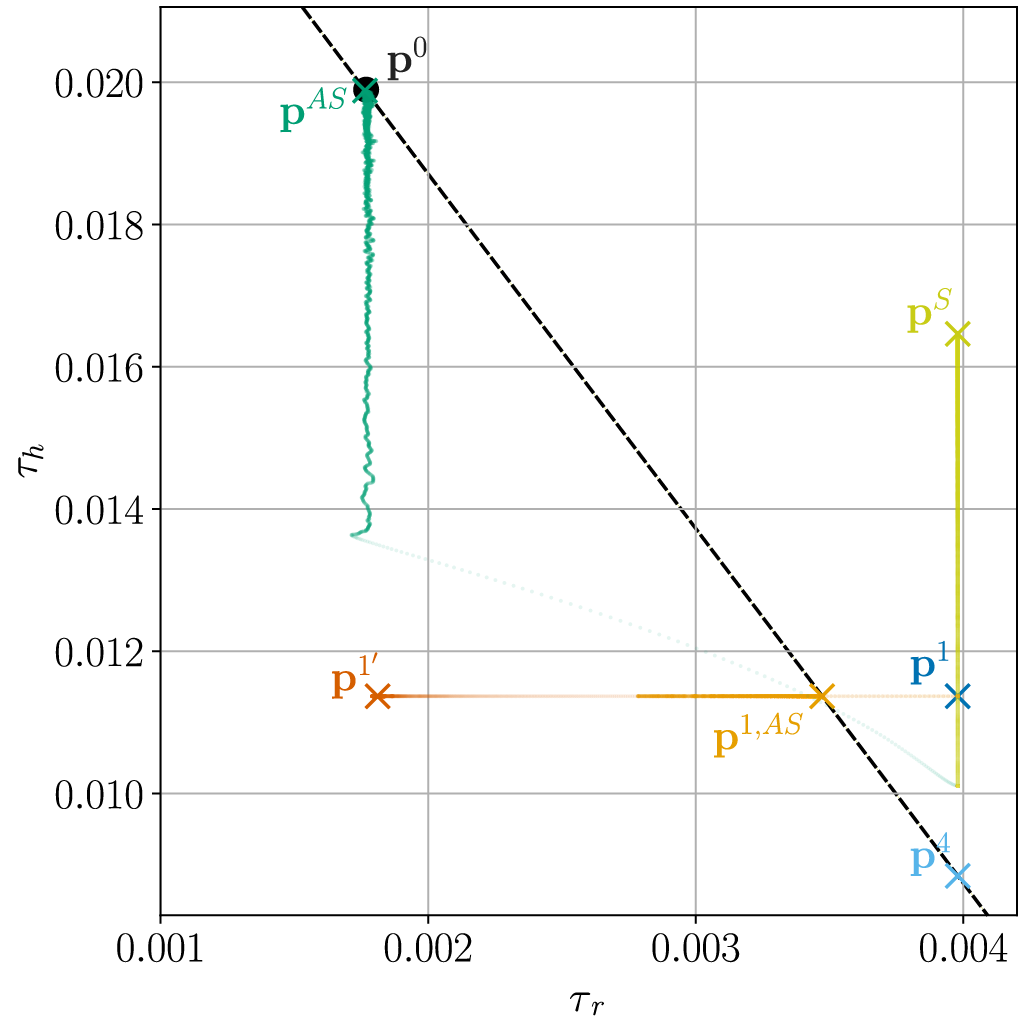}
        \caption{The values of $\tau_r$ and $\tau_h$ of model 1 ($p^1$, dark blue), model 4 ($p^4$, light blue), model 1' ($p^{1'}$, dark orange), model 1 trained as a member of ASUMO$_{14}$ ($p^{1,AS}$, orange), and the truth ($p^0$, large black dot) are shown. For SUMO$_{14}$ ($p^S$) and ASUMO$_{14}$ ($p^{AS}$), the effective parameter values are shown. The small dots indicate the values at each timestep during training, and the crosses mark the final values. The dashed line indicates the range of effective parameter values of ASUMO$_{14}$ when varying $w$.}
        \label{fig:parameter_space}
    \end{subfigure}%
    ~~~~~~
    \begin{subfigure}[t]{0.53\linewidth}
        \centering
         \includegraphics[width=\linewidth]{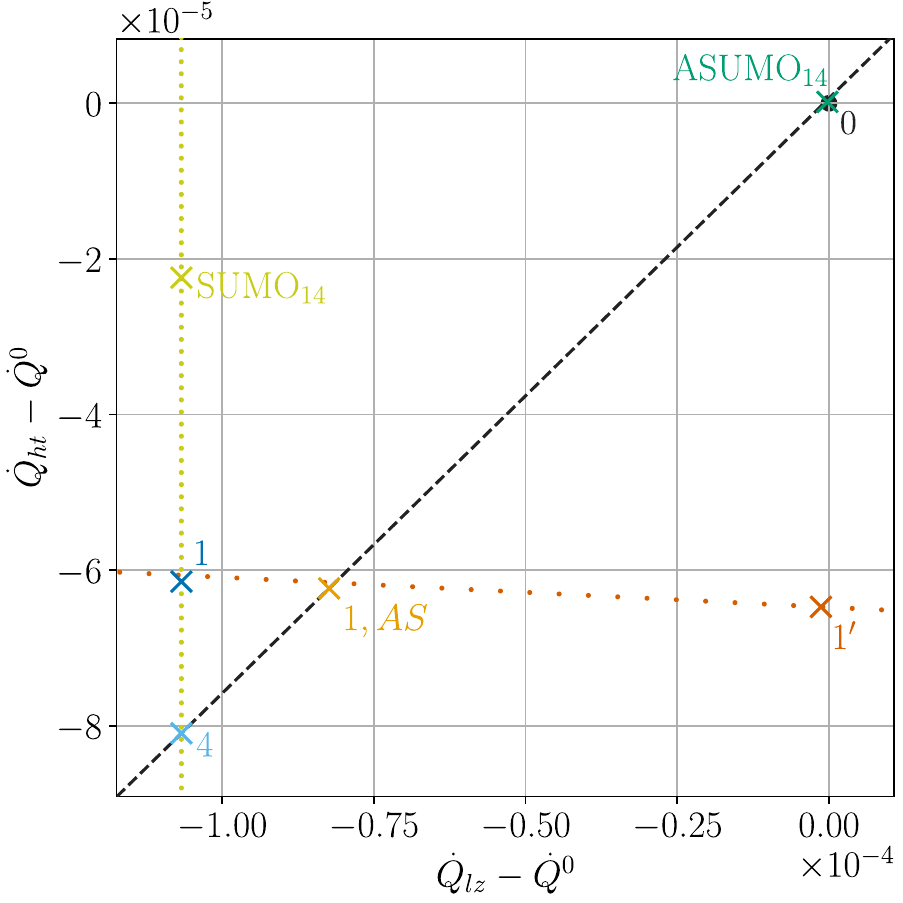}
        \caption{Error in the single-step tendency of a low zonal harmonic $\dot{Q}_{lz}$ (m=0, n=3) and a high tesseral harmonic $\dot{Q}_{ht}$ (m=10, n=21) of the potential vorticity. Crosses mark the errors of model 1, 4, 1', SUMO$_{14}$, ASUMO$_{14}$, and model 1 trained as a member of ASUMO$_{14}$ when run by itself (1,AS). The error of the perfect model 0 is highlighted as a large black dot at the origin. 
        Small dots in dark orange (yellow) indicate the tendency errors of model 1 (SUMO$_{14}$) for varied $\tau_r$ ($w$). The dashed line highlights the range of tendency errors of the ASUMO when varying $w$. }
        \label{fig:phase_space}
    \end{subfigure}
    \caption{Parameter and tendency space of the different methods used in \cref{sec:setup_asumo}.}
\end{figure*}

The parameter space of $\tau_r$ and $\tau_h$ of all models during training is shown in \cref{fig:parameter_space}. One can get closer to $\p^0$ by modifying $\tau_r$ of $\p^1$ ($\p^{1}{'}$), or using a weighted sum of $\tau_h$ of $\p^1$ and $\p^4$ (SUMO$_{14}$), but neither result in a good approximation of $\p^0$. With both degrees of freedom, an appropriate value for $\tau_r$ for one of the member models of ASUMO$_{14}$ ($\p^{1,AS}$) is found, making $\p^0$ accessible by an adequate value of $w$ ($\p^{AS}$). The magnitude of errors in parameters roughly correspond to the error in the state, see the tendency of two spherical harmonics of the models after training shown in \cref{fig:phase_space}. 

\begin{figure*}
    \centering
    \includegraphics[width=0.49\linewidth]{plots/expreview5_2A/mean_diff_earth_2_l2.pdf}
    \includegraphics[width=0.49\linewidth]{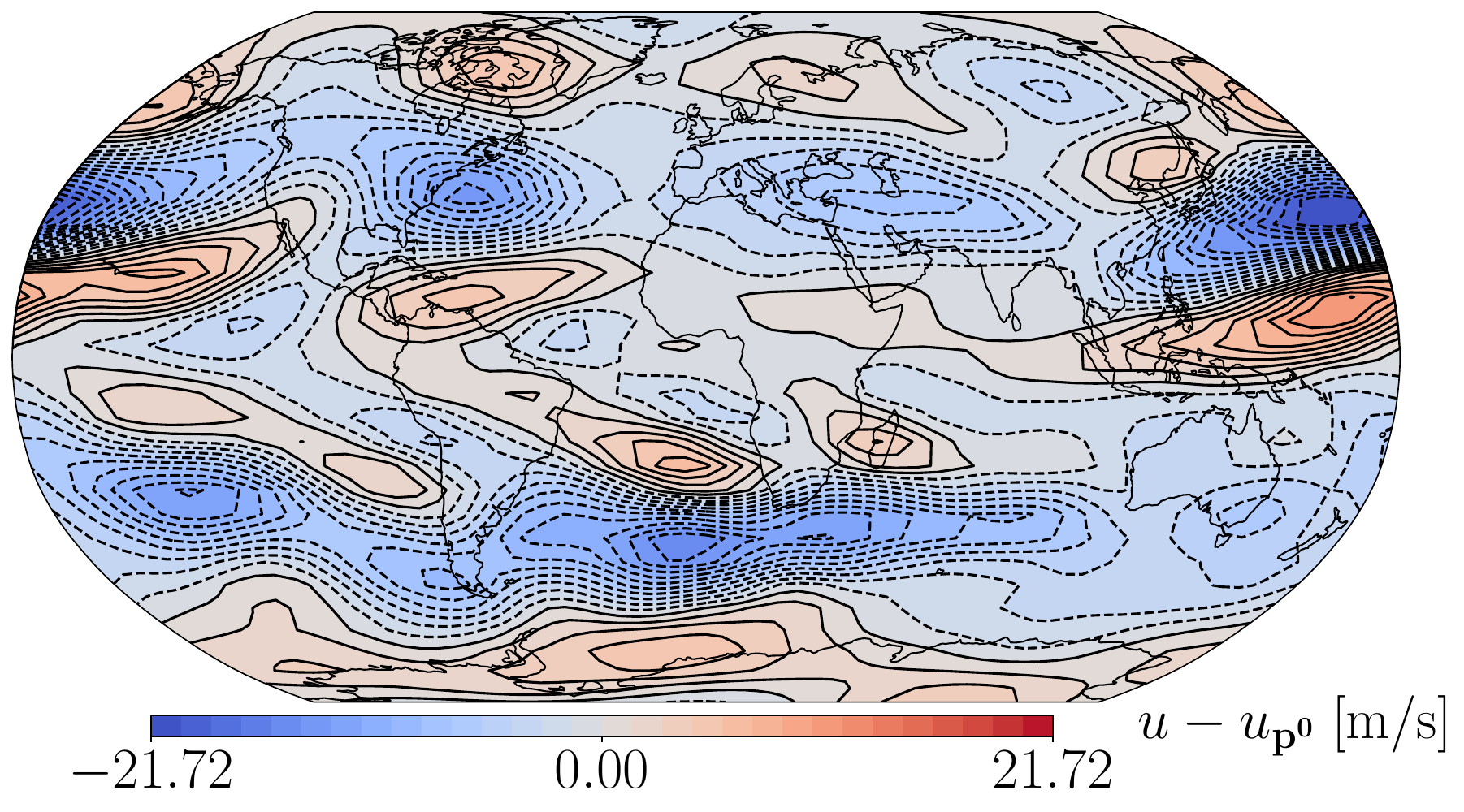}
    \includegraphics[width=0.49\linewidth]{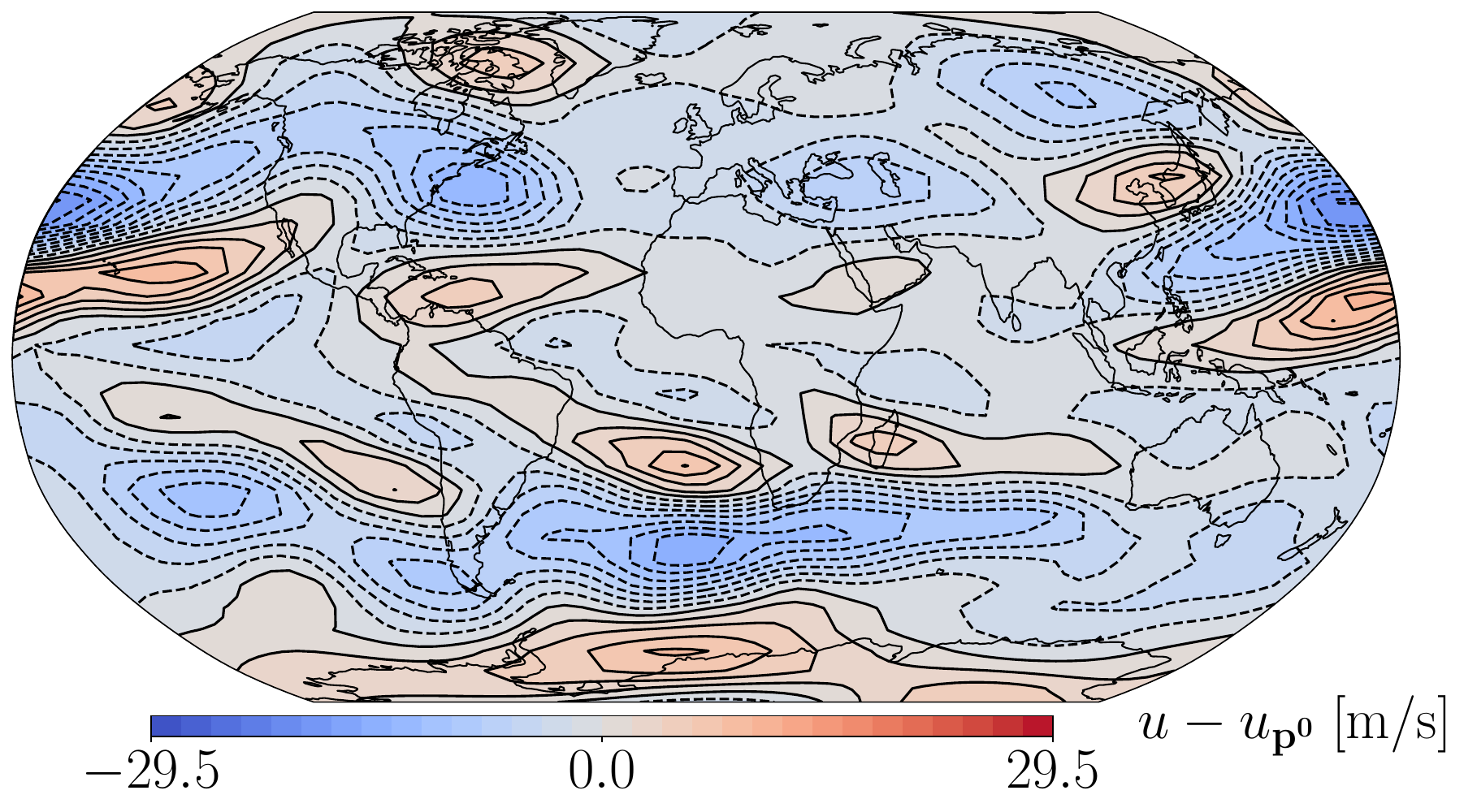}
    \includegraphics[width=0.49\linewidth]{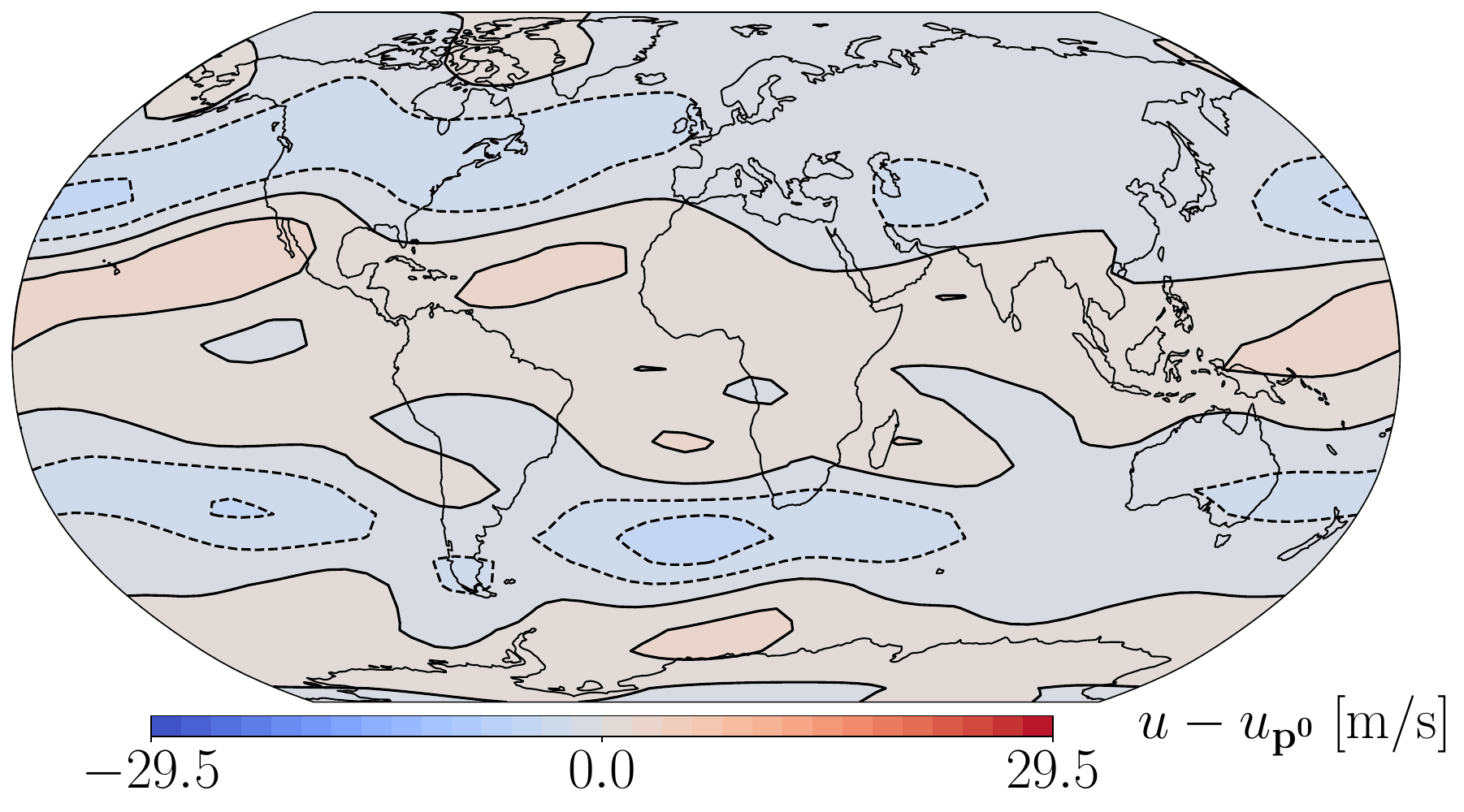}
    \includegraphics[width=0.49\linewidth]{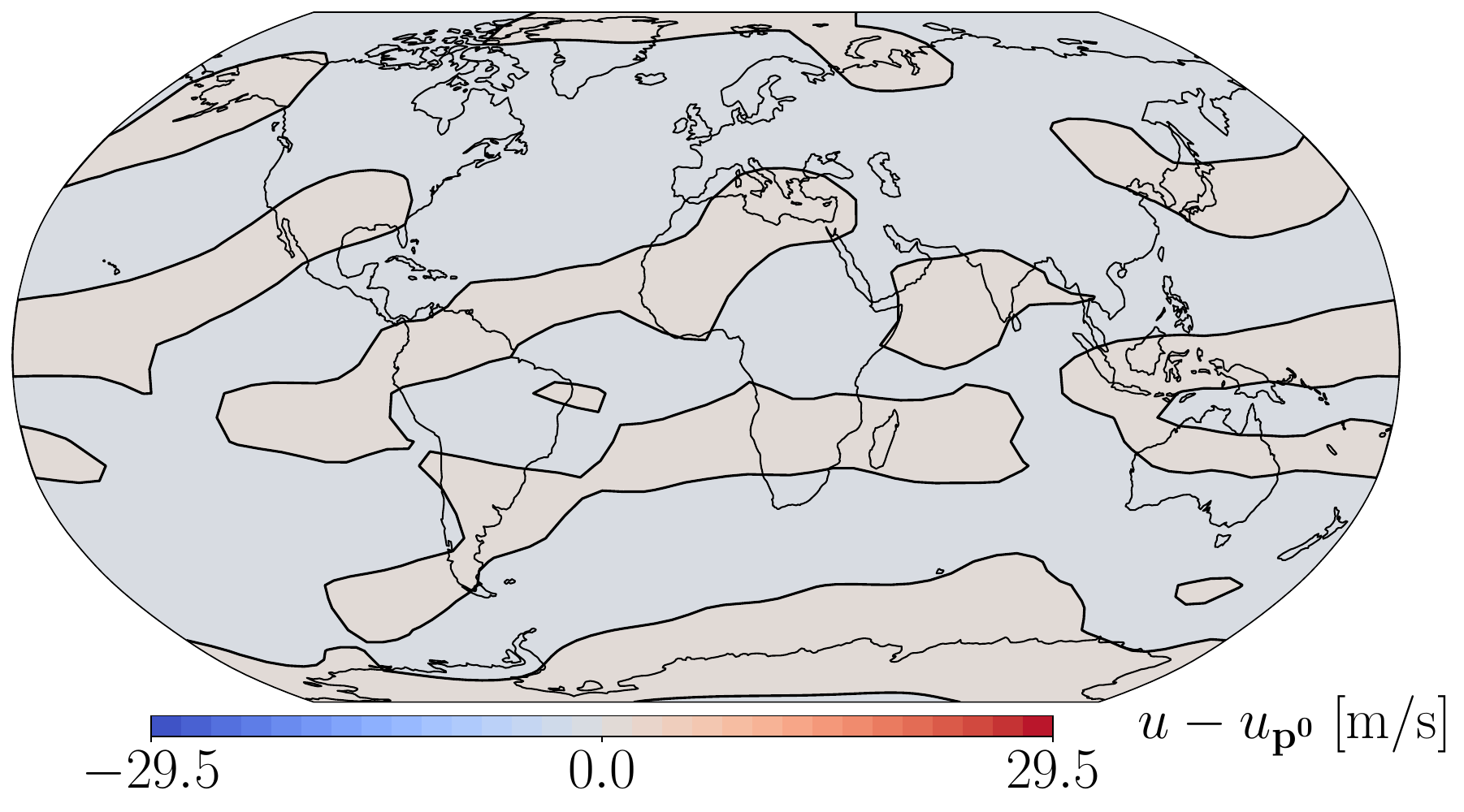}
    \caption{ Error in the climatological zonal winds at the 500 hPa level of, going left to right, from the top: model 1, model 4, SUMO$_{14}$, trained model 1, ASUMO$_{14}$. }
    \label{fig:earth_plots}
\end{figure*}

ASUMO$_{14}$ achieves the best climatological predictions, which are not significantly different from those of the perfect model, see the RMSEs of the zonal winds at the 500 hPa level for each experiment in \cref{tab:results_mean}. Further, note that despite a correct estimate of $\p^0$ being inaccessible to either training of parameters alone or SUMO by construction, both method still yield an improved performance in the climatology. A significant improvement is achieved by model $1'$ over model $1$ or $4$ by themselves, and a worse spread is traded off for a slightly improved mean in the case of SUMO$_{12}$. Climatological zonal winds are shown for every model configurations in \cref{fig:earth_plots}.
\begin{table}[hbtp]
\begin{center}
\begin{tabular}{p{7em}|ll}
\toprule
 & RMSE $\left<u\right>$ & RMSE $\sigma_u$ \\
\midrule
Model 0 & $0.53 \pm 0.04$ & $0.219 \pm 0.006$ \\
Model 1 & $4.526 \pm 0.017$ & $1.629 \pm 0.015$ \\
Model 4 & $4.653 \pm 0.018$ & $1.313 \pm 0.009$ \\
Model $1'$ & $1.54 \pm 0.04$ & $0.952 \pm 0.007$ \\
SUMO$_{14}$ & $4.55 \pm 0.03$ & $2.122 \pm 0.017$ \\
ASUMO$_{14}$ & $0.50 \pm 0.06$ & $0.240 \pm 0.006$\\
\bottomrule
\end{tabular}
\caption{ Root mean squared error of the temporal mean and standard deviation of zonal winds of all cells at the 500 hPa level with respect to the truth. }
\label{tab:results_mean}
\end{center}
\end{table}

\section{Discussion} \label{sec:discussion}

Three different model tuning methods were successfully applied to instantiations of a 3-level global atmospheric model with 1449 degrees of freedom. A gradient descent algorithm was first used to optimize 6 parameters of the model where all parameters were underestimated by between 20\% and 80\%, stabilizing to an error of $<20\%$ after training for 1500 days. This model achieved a global mean RMSE in zonal winds close to that of a perfect model. Then, the weight of a SUMO consisting of two imperfect models with non-trainable parameters was optimized, stabilizing after 250 days, and achieving a global mean RMSE of $<1.5$ m/s, compared to $>4.5$ m/s for any member model. This included a case where the SUMO members had the same sign parameter biases, which was optimized by finding a negative value for the weight.

Then, a new method, \emph{adapative supermodeling} (ASUMO) was introduced, which combines the training of both the weight of a SUMO and the internal parameters of one of its members. To display the advantage of ASUMO, two imperfect models were used where only some parameters were trainable and some parameters are identically biased to prevent (effectively) estimating the parameters of the truth by using either of the two first methods. All three methods were tested and compared. In all cases, the use of gradient descent resulted in some improvement in climatological forecasting skill over the starting model. The ASUMO reached the best approximation of the truth parameters, with a global RMSE within $8\%$ of that of the perfect model for the temporal mean, and $5\%$ for the standard deviation. 

These results demonstrate the robust performance of gradient descent algorithms in combination with supermodeling in a context where synchronization occurs, in face of noisy data from a climate model with a high number of degrees of freedom. They also show the potential of tuning long timescale performance by training parameters on short timescale processes. While we used the gradient descent algorithm here for demonstration purposes, other optimization algorithms could be used in its place in ASUMO. The results presented here were achieved with a training time of ~4 years with $2.6\%$ computational cost overhead per parameter. This computational cost is similar to other methods used in operational models, meaning that using the gradient descent algorithm, or replacing them with current methods such as EPPES, to implement ASUMO in operational models are both viable options.

The results presented in this paper are based on ideal situations: a ``truth" whose parameters can be exactly modeled, unbiased and uncorrelated noise, and the modest complexity of the model compared to modern climate models in operational use. Nonetheless, the good performance achieved at this stage warrants next-stage tests of ASUMO on more complex dynamical models and on real data. 




\begin{appendix}
\section{Derivation of Marshall-Molteni Model}\label{sec:qg}
Conservation of momentum on a rotating sphere and the first law of thermodynamics result in the standard filtered partial differential equations describing the temporal evolution of vorticity and temperature suitable to study the dynamics of atmospheric flow at mid-latitudes
\begin{align}
\frac{\partial}{\partial t} \zeta &= -\bv_{\psi} \cdot \nabla(\zeta + f) + f_0\frac{\partial \omega}{\partial p}\\
\frac{\partial}{\partial t} \frac{\partial \Phi}{\partial p} &= -\bv_{\psi} \cdot \nabla \frac{\partial\Phi}{\partial p} -\sigma \omega.
\end{align}
Relative vorticity is defined as the rotation of the horizontal wind, $\zeta \equiv \bk \cdot \nabla \times \bv$ with $\bv=(u,v)$, $u$ the east-west and $v$ the north-south component of the wind, $\nabla =\frac{1}{a}(\frac{1}{cos\phi}\frac{\partial}{\partial \lambda},\cos\phi\frac{\partial}{\partial \mu}$), $\lambda$ the geographic longitude and $\mu$ the sine of the geographic latitude $\phi$, $a$ the average radius of the Earth, $\bv_{\psi}$ the rotational part of the wind field that can be written in terms of the streamfunction $\psi$ as $\bv_{\psi}=\bk \times \nabla \psi$ with $\bk$ the vertical unit vector. The coriolis parameter $f=2\Omega \sin \phi$, with $\Omega$ the angular velocity of the earth, describes the contribution of the Earth's rotation to the vorticity of an air parcel at latitude $\phi$ and $f_0$ is its value at a particular reference latitude. It is also referred to as planetary vorticity. Pressure $p$ is used as a vertical coordinate, $\omega$ is the pressure velocity which is defined as the Lagrangian rate of change of pressure with time. The vorticity equation states that the local rate of change of vorticity is due the horizontal advection of relative and planetary vorticity plus the generation of vorticity due to vertical stretching. Forcing and dissipation terms have been omitted for simplicity. Temperature is written as the pressure derivative of the geopotential $\Phi$ under the assumption of hydrostatic balance  and application of the ideal gas law.
Hydrostatic balance states that the pressure is equal to the weight of the atmospheric column above
\begin{equation}
\mathrm{d}p=-\rho g \mathrm{d}z \equiv -\rho \mathrm{d} \Phi,
\end{equation}
where $g$ denotes the gravitational acceleration, $\rho$ the density of air and $z$ the height. Application of the ideal gas law $p=\rho RT$ gives a relation between temperature and geopotential
\begin{equation}
\frac{\partial\Phi}{\partial p}=-\frac{RT}{p},
\end{equation}
with $R$ the gas constant. Finally $\sigma$ in the temperature equation denotes the vertical stability. The temperature equation states that the local rate of change of temperature is due to the horizontal advection of temperature and adiabatic heating due to vertical displacements. Combination of the vorticity and temperature equation and using the approximate linear balance equation $\nabla \Phi = f_0 \nabla \psi$ leads to a single equation for a quantity called potential vorticity (PV) that is conserved following the motion in the absence of forcing and dissipation
\begin{equation}
\left(\frac{\partial}{\partial t}+\bv_{\psi} \cdot \nabla \right) \left(  \zeta + f + f^{2}_0\frac{\partial}{\partial p}\sigma^{-1}\frac{\partial \Phi}{\partial p}\right)=0.
\end{equation}
The quasi-geostropic model solves this partial differential equation in a finite state space with vorticity defined at discrete pressure levels 200 (level 1), 500 (level 2)  and the 800 hPa level (level 3) and temperature at 650 and 350 hPa
\begin{align}
\frac{\partial q_1}{\partial t} &= -\bv_{\psi_1} \cdot \nabla q_1  - D_1(\psi_1,\psi_{2}) + S_1\nonumber \\
\frac{\partial q_{2}}{\partial t} &= -\bv_{\psi_2} \cdot\nabla q_{2}  - D_{2}(\psi_1,\psi_{2},\psi_{3}) + S_{2}\nonumber \\
\frac{\partial q_{3}}{\partial t} &= -\bv_{\psi_3} \cdot \nabla q_{3}  - D_{3}(\psi_{2},\psi_{3}) + S_{3},
\label{eq:t213l}
\end{align}
where $q$ is PV, $D(\psi)$ is a linear operator representing dissipative terms, $S$ is a constant PV forcing, and the index $i=1,2,3$ refers to the pressure level. Here, PV is defined as
\begin{align}
q_1 &= \nabla^{2}\psi_1 - R_1^{-2}(\psi_1-\psi_{2}) + f \nonumber \\
q_{2} &= \nabla^{2}\psi_{2} +R_1^{-2}(\psi_1-\psi_{2})  - R_{2}^{-2}(\psi_{2}-\psi_{3})+ f \nonumber \\
q_{3} &= \nabla^{2}\psi_{3} + R_{2}^{-2}(\psi_{2}-\psi_{3}) + f(1+\frac{h}{h_0}),
\end{align}
where $R_1$ (=700 km) and $R_{2}$ (=450 km) are Rossby radii of deformation appropriate to the 200-500 hPa layer and the 500-800 hPa layer, respectively and $h_0$ is a scale height set to 3000 m and $h$ the height of the topography. The topography term has entered the equation through the lower boundary condition where flow over mountains leads to vertical displacements of air and the generation of vorticity through stretching.
In the horizontal the equations are solved by a Galerkin projection of (\cref{eq:t213l}) onto a basis of spherical harmonics 
\begin{equation}
Y_{m,n}(\lambda,\mu)=P_{m,n}(\mu)e^{im\lambda},
\end{equation}
where $P_{m,n}(\mu)$ denote associated Legendre polynomials of the first kind, $m$ the zonal wavenumber and $n$ the total wavenumber. The spherical harmonics
are eigenfunctions of the Laplace operator: 
\begin{equation} 
\Delta Y_{m,n}(\lambda,\mu) =
-n(n+1)Y_{m,n}(\lambda, \mu), 
\end{equation}
and obey the following orthogonality condition
\begin{align}
\left<Y_{m,n},Y_{m',n'}\right>\,=\,&\frac{1}{4\pi}\int_{o}^{2\pi}\int_{-1}^1 Y_{m,n},Y_{m',n'}\,d{\lambda}d{\mu}\nonumber\\
=&\delta_{mn,m'n'}.
\end{align}

A triangular truncation of this expansion at total wavenumber 21 ($0<n\leq21, -n \geq m \leq n$) leads to a system of 1449 coupled ordinary differential equations for the coefficients $Q_{lmn}$ of the spherical harmonical functions\footnote{The complex coefficients $Q_{lmn}$ are real-valued for the zonal components ($m=0$) and the wave components ($m \neq 0$) obey $\operatorname{Im}(Q_{lmn})=-\operatorname{Im}(Q_{l-mn})$.}, 483 at each of the three pressure levels
\begin{align}
q_{l}(\lambda,\mu,t)=\sum_{n=1}^{21}\sum_{m=-n}^{n} Q_{lmn}(t) Y_{m,n}(\lambda,\mu) \nonumber\\
\hspace{3cm}\mbox{for}\quad l=1,2,3.
\end{align}

We then write the set of coupled ODEs as
\begin{equation}
\dot{\Q} = \F(\Q,\p),
\end{equation}
where $\Q\equiv(Q_{000}, Q_{001}, \ldots,Q_{lmn})$ denotes the vector of PV expansion coefficients at every level $l$, and wavenumbers $m$ and $n$, and $\F\equiv(F_{000}, F_{001}, \ldots,F_{lmn})$ is a vector function returning the tendencies of each $Q_{lmn}$, which depends on $\Q$ and on parameter vector $\p$.

In (\cref{eq:t213l}), $D_1, D_{2}, D_{3}$ are linear operators representing the effects of Newtonian relaxation of temperature ($\pazocal{R}$), Ekman dissipation of  vorticity due to linear drag on the 800 hPa wind ($\pazocal{E}$), and horizontal diffusion of vorticity ($\pazocal{D}$)
\begin{align}
-D_1 &= \pazocal{R}_{12} - \pazocal{D}_1\\
-D_2 &= -\pazocal{R}_{12} + \pazocal{R}_{23} - \pazocal{D}_2\\
-D_3 &= -\pazocal{R}_{23} - \pazocal{E}_3 - \pazocal{D}_3.
\end{align}
The term
\begin{equation}
\pazocal{R}_{12} = \tau^{-1}_{R}R^{-2}_1(\psi_1-\psi_{2})
\end{equation}
describes the effect of temperature relaxation between levels 1 and 2 due to radiative cooling, with a radiative time scale $\tau_{r}=20$ days; the corresponding term for temperature relaxation between levels 2 and 3 is
\begin{equation}
\pazocal{R}_{23} = \tau^{-1}_{R}R^{-2}_{2}(\psi_{2}-\psi_{3}).
\end{equation}
The Ekman dissipation is given by
\begin{equation}
\pazocal{E}_{3} = \bk \cdot \nabla \times [c_d(\lambda,\phi)\bv_{\psi_3}].
\end{equation}
The drag coefficient $c_d$ is dependent on the land-sea mask and the orographic height
\begin{equation}
c_d(\lambda,\phi)=\tau^{-1}_{E}[1+\alpha_1M(\lambda,\phi)+\alpha_{2}H_d(\lambda,\phi)]
\end{equation}
with the time scale of the Ekman damping $\tau_{E}=3$ days, $\alpha_1 = \alpha_2 = 0.5$; $M(\lambda ,\phi)$ is the fraction of land within a grid box; and
\begin{equation}
H_d(\lambda,\phi)=1- e^{-\frac{h(\lambda,\phi)}{1000}}.
\end{equation}
Since $M$ and $H_d$ vary between 0 and 1, $c_d$ varies between (3 days)$^{-1}$ over the oceans, (2 days)$^{-1}$ over zero altitude land and about (1.5 days)$^{-1}$ over mountains higher than 2000 m. Finally, at each pressure level $l$, the time-dependent component of PV $q_l'$ (i.e. PV minus planetary vorticity and orographic component) is subject to a scale-selective horizontal diffusion
\begin{equation}
\pazocal{D}_l = c_h\nabla^{p_h}q_l',
\end{equation}
where the coefficient
\begin{equation}
c_h=\tau_h^{-1}a^{p_h}(21\cdot 22)^{-\frac{p_h}{2}}.
\end{equation}
With the power $p_h$ set to 4, $c_h$ is such that spherical harmonics of total wavenumber 21 are damped with time scale $\tau_h=2$ days.
The PV source terms $S_{i}$ in (\cref{eq:t213l}) are calculated from observations as the opposite of the time-mean PV tendencies obtained by inserting observed daily winter time stream function fields into (\cref{eq:t213l}) with the PV source terms set to zero.

\section{Robustness}\label{sec:robustness}
\begin{figure}
    \centering
    \includegraphics[width=\linewidth,trim={0 0 0 0em},clip]{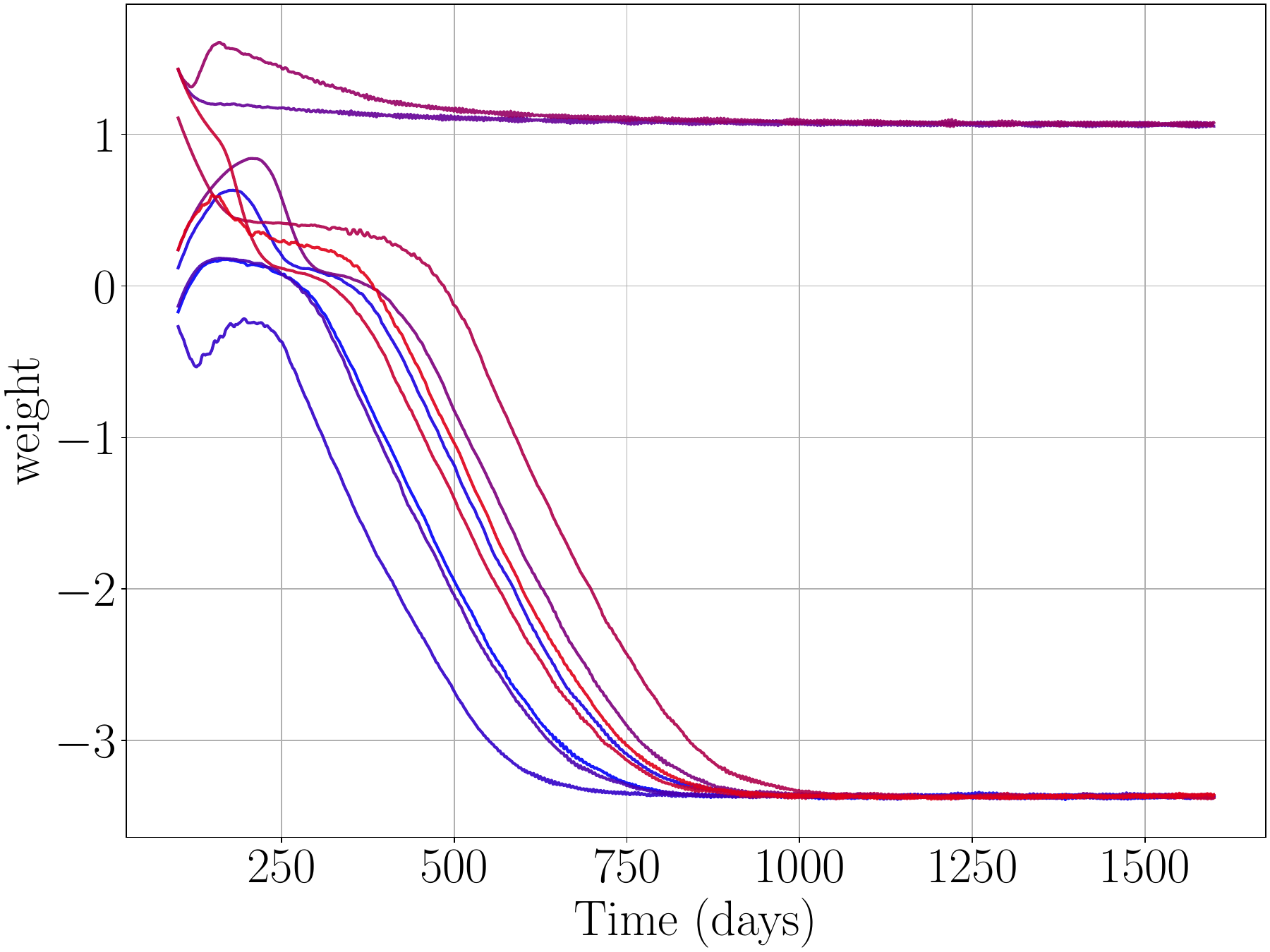}
    \includegraphics[width=\linewidth,trim={0 0 0 0em},clip]{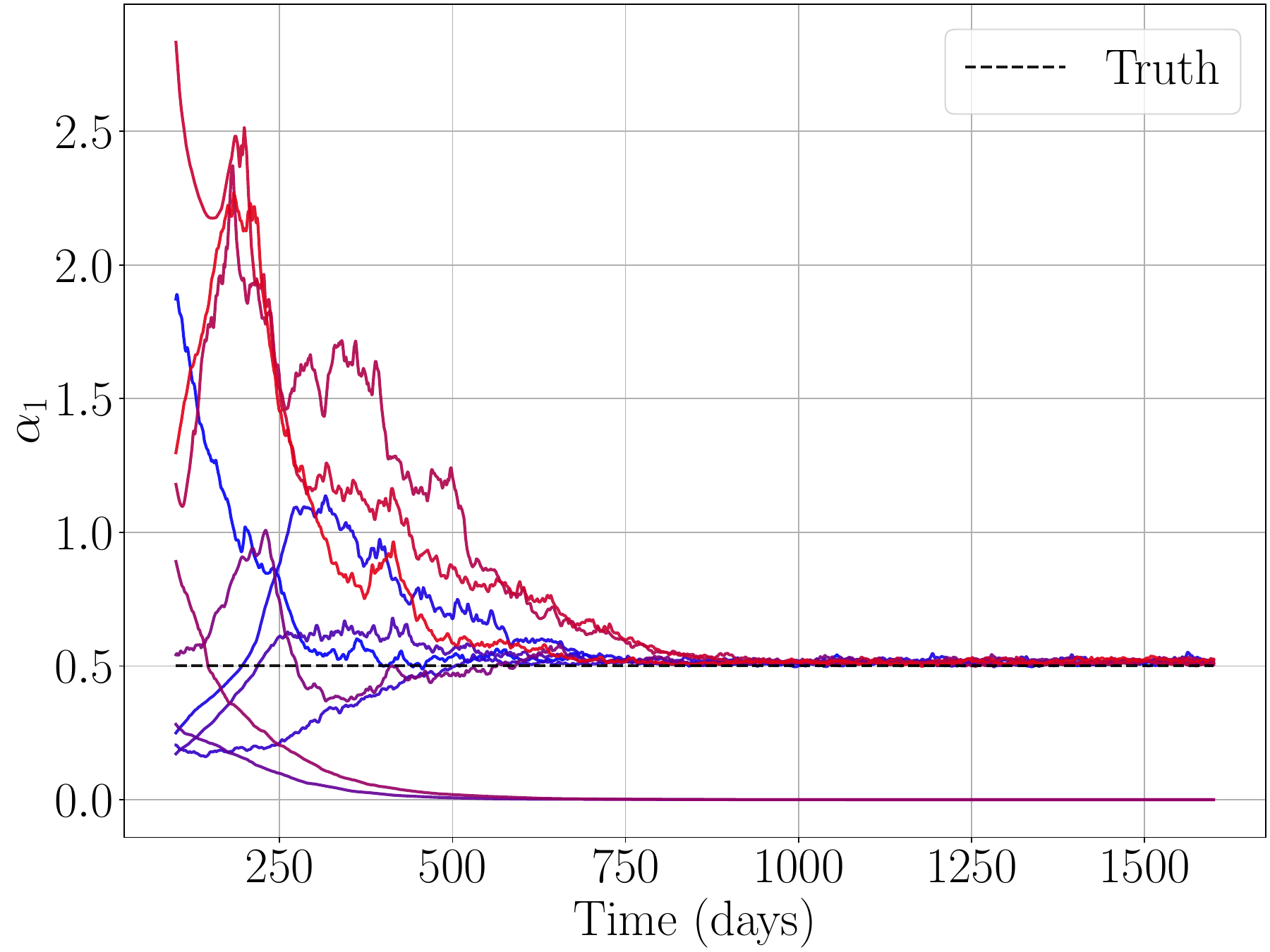}
    \caption{Training of weight (top) and $\alpha_1$ (friction over land, bottom) for ten experiments all starting with scrambled starting parameter values. In 2 of the cases, the optimization fails. Otherwise the experiments converge to similar configurations. }
    \label{fig:local_minima_addish}
\end{figure}

To test the robustness of ASUMO against starting values and local minima, the ASUMO experiment described in \cref{sec:opt_parameters} was repeated 10 times with randomized starting values ranging from 10\% to 10 times the truth values. 8 of the experiments converged to solutions with parameter values close to those found in \cref{sec:asumo}. The friction over topography showed the largest spread, with a mean value of $0.51$ and a standard deviation of $0.007 (1.4\%)$, shown in \cref{fig:local_minima_addish}. In 2 of the experiments, the optimization failed due to the vanishing of the friction over land and topography, and the temperature relaxation. Our implementation of the algorithm is agnostic to the physics of the model, and we only prevent parameters from adopting negative values. In practice, the vanishing of a parameter can be remedied with conservative constraints on parameter values.

\section{Computational Cost}\label{sec:cost}
Our implementation of the update rule used analytically calculated gradients and consumed 2.6\% more computational power compared to running the model alone per internal parameter, see \cref{tab:parameters_computational}. Compared to an operational model, our model is relatively light and has few functions, meaning that each operation on a parameter will represent a larger fraction of CPU compared to an operational model.

\begin{table}[hbtp]
\begin{center}
\begin{tabular}{p{3.3 cm} |l }
Call to function & CPU time [ms]  \\
 \hline
 \hline
 Model & $1.793 \pm 0.011$ \\
\hline
Update rule 1 parameter (est.)& $0.0463 \pm 0.0003$ \\
\hline
Update rule 6 parameters & $0.278 \pm 0.005$ \\
\end{tabular}
\caption{ Computational cost of one iteration of the model, update rule for one parameter, and six parameters.}
\label{tab:parameters_computational}
\end{center}
\end{table}

The computational cost of implementing ASUMO is dependent on its implementation. Using a first-order finite-difference method for the parameter optimization would come at computational complexity of $\sim O(n)$ where $n$ is the number of parameters being estimated. Analytically solving the gradients however would result in a much more efficient implementation, but is often difficult to implement in modern operational models. In the case of the supermodel, the computing costs for running each model must be added, compared to a negligible costs of updating the weights, resulting in complexity $O(m)$ where $m$ is the number of model members in the SUMO.

State of the art methods such as EPPES typically perform 10 years of integration to estimate 20 parameters. Taking the overhead and shorter timesteps into account (40 minutes compared to 60 minutes for IFS), we estimate a similar computational cost of our method per parameter, per member model. Implementing ASUMO with EPPES or other tuning algorithms would then come at a cost of $C_\text{ASUMO}=mC$ where $m$ is the number of member models in the ASUMO and $C$ is the current computational cost for using EPPES on a single model.

\section{Adaptive Learning Rates with Adam Optimizer}\label{sec:adam}

Different parameters will train at speeds that can differ by multiple orders of magnitude. For this reason, the Adam (Adaptive Moment Estimation) optimizer \cite{adam} is implemented here to automatically adjust the rates of learning of each parameter. Adam works by continuously estimating the mean ($p_t$) and variance ($q_t$) of the gradients of the function to optimize. In our case, we use the update rule $U$ as the gradient. For timestep $i$,
\begin{align}
    \hat{p}_t &= \frac{p_t^i}{1-m_1^{i}} \\
    \hat{q}_t &= \frac{p_t^i}{1-m_2^{i}}
\end{align}
where $p_t^i = m_1p_t^{i-1}+(1-m_1)U$ and $q_t^i = m_2q_t^{i-1}+(1-m_1)U^2$, and $ m_1 = 0.9$ and $m_2=0.9999$ are standard decay factors for the moving averages of the gradient and squared gradient. The gradient is updated by
\begin{equation}
\text{Adam}(U) = \alpha \frac{\hat{p}_t}{\sqrt{\hat{q}_t} + \epsilon}    
\end{equation}
where $\alpha=0.001$ is a standard learning rate and $\epsilon=10^{-8}$ avoids division by zero.

\end{appendix}

\newpage
\twocolumn[
\begin{center}
\addcontentsline{toc}{section}{References}
\end{center}
]


\printbibliography

\end{document}